\documentclass[fleqn,usenatbib]{mnras}

\usepackage{newtxtext,newtxmath}

\usepackage[T1]{fontenc}

\DeclareRobustCommand{\VAN}[3]{#2}
\let\VANthebibliography\thebibliography
\def\thebibliography{\DeclareRobustCommand{\VAN}[3]{##3}\VANthebibliography}

\usepackage{graphicx}	
\usepackage{amsmath}	

\usepackage{float}

\title[Vertical DFs for the MW]{Non-isothermal vertical distribution functions for the Milky Way}

\author[M. Djurić et al.]{
Maria Djurić$^{1}$\thanks{E-mail: maria.djuric.22@ucl.ac.uk} and 
Ralph Sch\"{o}nrich$^{1}$
\\
$^{1}$Mullard Space Science Laboratory, University College London, Holmbury St. Mary, Dorking, Surrey RH5 6NT, UK\\
}

\date{Accepted XXX. Received YYY; in original form ZZZ}

\pubyear{2026}

\begin{document}
\label{firstpage}
\pagerange{\pageref{firstpage}--\pageref{lastpage}}
\maketitle

\begin{abstract}

The vertical structure of our Galaxy has commonly been assumed to follow a pseudo-isothermal distribution. However, there is no \textit{a priori} reason to expect this form to arise from scattering by giant molecular clouds (GMCs), since GMCs are confined to a narrow layer around the Galactic midplane and therefore do not randomise stars uniformly in vertical phase space. We first present a simple statistical argument to derive a new vertical distribution function (DF), in the limiting case of a razor-thin distribution of GMCs, which admits an extra factor of ${J_z}^{-\alpha}$, where $\alpha>0$ and varies for different potentials. In light of this, we revisit the diffusion coefficients for vertical heating from the mechanics of two-body scattering, in a more realistic, extended distribution of GMCs. The resulting drift and diffusion coefficients are strongly non-linear in $v_z$, particularly at low relative velocities, where the effective range of weak, local encounters can vanish. We determine stationary solutions, and also numerically solve the orbit-averaged Fokker--Planck equation with time-dependent coefficients. We find, in both cases, a vertical action distribution that deviates from that of the pseudo-isothermal case, and propose a new, family of vertical DFs that generalise the pseudo-isothermal form. We find the two primary agents of non-isothermality are: (i) a GMC distribution confined near the midplane, and (ii) the fact that the Coulomb logarithm, $\Lambda$ is strongly velocity-dependent at small velocities, and cannot be treated as a constant, contrary to the commonly adopted assumption. \\
\end{abstract}




\section{Introduction}
\par
Among the most central, unresolved problems in the evolution of galactic discs is a precise description of stellar kinematics, namely, the motion of stars predicted as a consequence of disc evolution. In a simple picture, we can separate the disc into different stellar populations that are born from the interstellar medium, so each starts on nearly circular orbits with some (typically) minor random motions. It was \citet{Spitzer_1942} to suggest  the use of the isothermal-sheet distribution function (DF) for the \textit{vertical} component, $f_z(E_z)\propto \exp(-E_z/\sigma_z^2)$, equivalent to a Gaussian in $v_z$. For stellar discs, \citet{Shu_1969} then derived the widely adopted Shu DF, a modification of the Schwarzschild DF, which also corresponds to a Maxwellian velocity distribution. It has also empirically been found that mono-abundance and mono-age populations (e.g. \citealt{Bovy_2012}), also exhibit near-isothermal kinematics. Assuming separability of vertical from in-plane components, \citep{Carlberg_Sellwood_1985}, the isothermal prescription can be recast in action space, by using an epicycle approximation, in which $E_z \simeq \Omega_z \, J_z$, and thus the Maxwellian velocity distributions are transcribed to an exponential action distribution (e.g \citet{Binney_2010}),
\begin{equation}
    f_z(J_z) \propto e^{-\Omega_z J_z/\sigma_z^2}.
\end{equation}
This line of reasoning has supported the widely used family of quasi-isothermal, action-based disc DFs employed in dynamical modelling of the Milky Way (MW) (e.g. \citealt{Binney_2012,Bovy_Rix_2013}).
\par
A key problem with adopting such DFs is that it is not the result of a rigorous relaxation argument, but is rather a convenient assumption from the idea of many small, uncorrelated scatterings leading to an approximately Gaussian velocity distribution. However, in realistic discs, strong anisotropy of the velocity ellipsoid ($\sigma_R > \sigma_z$) \citep{Siebert_2011},
the co-existence of different populations with different kinematics, and the extremely long relaxation times indicate that simply assuming a Maxwellian is not warranted (e.g. \citealt{Spitzer_Schwarzschild_1951,Dehnen_Binney_1998,Sellwood_2014}). Moreover, \textit{Gaia} \citep{DR3} has revealed clear signatures of non-equilibrium in the solar neighbourhood, most notably the vertical phase-space spiral \citep{Antoja_2018}, which is interpreted as ongoing phase mixing following a perturbation  \citep{BinneySchonrich2018,JBH2021}. Despite the problem with assuming Gaussian velocity distributions, there has been a long tradition in the field of addressing kinematics of different populations and disc heating in the language of velocity dispersions, $\sigma_i$, and variation with height above the plane, as a function of age, (e.g. \citealt{Dehnen_Binney_1998, Nordstrom_2004,Aumer_Binney_2009,ABS_2016}) and/or metallicity (e.g. \citealt{Bovy_2012,Liu_Glenn_2012}). Harnessing the power of multiple high-precision spectroscopic and photometric surveys, it has recently become possible to investigate both simultaneously (e.g. \citealt{Mackereth_2019}). Similarly, a longer tradition of papers have employed analytical models \citep{Spitzer_Schwarzschild_1951,Lacey_1984,Binney_Lacey_1988,Jenkins_Binney_1990} and simulations \citep{Hannien_Flynn_2002,ABS_2016} to predict the velocity dispersions of stellar populations from different processes. These studies have also identified the main sources of heating, notably, the role of giant molecular clouds (GMCs) in heating the vertical component of our Galaxy. To investigate disc heating, one can begin with an adequate equilibrium DF for the vertical component of our Galaxy, which evolves secularly from internal or external perturbations on the Galaxy. On the assumption that GMCs are the primary source of \textit{vertical} heating \citep{Ida_1993,Sellwood_2014,ABS_2016,Mackereth_2019}, it is natural to consider the role of such star-cloud interactions on the distribution of vertical velocities, and resultantly, the DF. 
\par
The commonly adopted isothermal sheet, for the vertical component \citep{Spitzer_1942}, assumes a constant stellar vertical velocity dispersion. Observations of composite populations, however, indicate that $\sigma_z$ varies with height above the plane, and in particular, have revealed linear velocity dispersion gradients with height  \citep{Jing_2016,Hagen_2018,Guo_2020}. Early investigations of a non-isothermal velocity dispersion were considered by \citet{Camm_1950} and \citet{Perry_1969}, although there has been little discussion on this, and the pseudo-isothermal case is most commonly adopted, largely as an approximation and mathematical convenience. 
\par
More recently, \citet{Sarkar_Jog_2018,Sarkar_Jog_2020} investigated the dynamical effect of a non-isothermal velocity dispersion, and had found, in the presence of a coupled system of stars, gas and a halo, that the midplane density is higher than that of the isothermal case. Analogously, \citet{Li_Widrow_2021} studied the MW's vertical distribution function using Gaia DR2 \citep{DR2} stars, and introduced an alternative distribution function. Again, the function is consistent with a non-isothermal velocity dispersion, and is steeper than the isothermal sheet at small actions, but is not consistently derived from a relaxation argument. In light of this, we motivate a revision of the isothermal assumption, and propose a model that is physically motivated by the mechanics of two-body scatterings and the spatial distribution of GMCs.
\par
This issue also matters for dynamical inference. Despite evidence for a non-isothermal vertical component, many classical determinations of the local vertical force and surface density assume an isothermal distribution for the tracer population. Neglecting velocity dispersion gradients in the Jeans equations can bias the inferred vertical forces, and thus determinations of dark matter and baryon densities. It has been shown that the disc cannot be modelled with a single DF, and requires a decomposition of components, in particular thick and thin disc stars \citep{Binney_2010}, or into chemically homogeneous sub-populations \citep{Bovy_2012}. The disc component of the MW is typically written as a superposition of isothermal (or pseudo-isothermal) components \citep{Flynn_2006}. Our aim in this paper, however, is to understand whether a physically motivated heating mechanism via GMC scattering modifies the shape of the vertical DF in action for coeval populations, away from the exponential form predicted by globally homogeneous diffusion.
\par
Moreover, since GMC density falls off rapidly with height  \citep{Villumsen_1985}, the scattering is no longer randomised in phase space but rather concentrated near the Galactic midplane, which immediately induces $z-$dependent scattering. In a recent study of small-scale scattering, such as that with GMCs, \citet{Tremaine_2023} performed a test-particle simulation to model these perturbations as a random walk process. Two simplifying assumptions here were: (i) assuming that the scattering was not concentrated in the Galactic midplane, and (ii) a symmetric random walk scattering model with a constant diffusion to model the vertical heating. Thus, the resulting equilibrium vertical action distribution was exponential.  Our approach differs in that we treat GMC scattering as explicitly dependent on phase-space coordinates, and, as we show, strongly dependent on velocity, which produces deviations from a pure exponential in $J_z$. This distinction is particularly relevant when modelling the diffusion of the vertical phase-space spiral, with lifetimes of $\lesssim 1$~Gyr, where the rate of phase mixing depends on how diffusion varies over the orbit. Similarly, \citet{Chiba_2025_2_spirals} implement small-scale kicks as in \citet{Tremaine_2023}, and argue the mean change in action is independent of the choice of coordinate system in which the scattering occurs. We believe this is only true if scattering probabilities are constant in all of phase space, and the diffusion coefficient of scattering is constant, which we will show in \S \ref{sect:lacey-scattering}, is not the case for a more realistic configuration.
\par
This paper is structured as follows. In \S \ref{sect:separable-df} we discuss the grounds on which the DF is separable, and state the commonly adopted vertical DF for our Galaxy. In \S \ref{sect: DF-GMC}, we argue a modified DF is required in the presence of star-cloud scattering with GMCs, in the limiting case of a razor-thin distribution. This is followed by an analysis of scattering in a realistic, extended distribution of GMCs, and a revision of diffusion coefficients in \S \ref{sect:lacey-scattering}. In \S \ref{sect:new-df}, we propose a modified DF for time-independent stationary solutions, and in \S \ref{sect:new-test-particle-sim}, we directly simulate the Fokker-Planck equation with time-dependent coefficients, and generalise the modified DF to vary with time. We discuss assumptions and limitations in \S \ref{sect:discussion}, and finally, in \S \ref{sect:conc}, we conclude and state final remarks. 

\section{A separable DF} \label{sect:separable-df}
\par
By Jeans' Theorem \citep{Jeans_1915}, for a system in near-equilibrium, the DF depends explicitly only on the integrals of motion, $I(\pmb{x},\pmb{v})$, which are constant along orbits. Any DF can be expressed as a DF in the integrals of motion \citep{Arnold_1978}, from which distributions in phase space (x,v) can be determined. In past decades, distributions primarily involving energy and angular momentum components have been popular, but more recently, these have been largely replaced by distributions in the three classical actions $\mathbf{J} = (J_R, J_\phi,J_z)$, which, in contrast to energy, have the advantage of being approximately stable (for non-resonant orbits) against adiabatic changes of the system, such as through mass growth of the Galactic disc. In the Solar neighbourhood, for stars with small radial excursions, the vertical component of the DF, $f_z$ can be decoupled from the in-plane component, and $f_z$ effectively becomes only a function of the vertical action, $J_z$ \citep{Carlberg_Sellwood_1985}, such that
\begin{equation}
    f(J_R,J_\phi, J_z) = f_p(J_R,J_\phi)f_z(J_z | J_R, J_\phi),
\end{equation}
where $f_p$ and $f_z$ are the (normalised) DFs for in-plane and vertical motion, respectively, and $J_R$ and $J_\phi$ are the radial and azimuthal action. We remark that a formulation in actions fully captures the coupling between horizontal and vertical components of motion of non-resonant stars, while \textit{this} formulation tacitly omits that stars in vertical mean-motion resonances can exchange these actions and thus modify the distribution \citep[see, e.g.][for a treatment of resonances]{Binney_2016}, or undergo resonant pumping \citep{Sridhar_Touma_1996}. The main finding of this work concerns the core of the distribution function at small vertical action. In this regime, where there are generally no nearby mean-motion resonances, both radial and vertical motions -- and their corresponding actions -- separate extremely well, as the potential is close to the epicyclic approximation. In this paper, we will omit the effects of such resonances and focus on the vertical component.

\subsection{The vertical component}
\par
The isothermal sheet $f_z \propto e^{-E_z/\sigma_z^2}$ \citep{Spitzer_1942}, where $\sigma_z$ is the vertical velocity dispersion, is a common choice of DF for the vertical profile of our Galaxy. To write the DF in action space, \cite{Carlberg_Sellwood_1985} and later, \cite{Binney_2010}, employed the epicycle approximation, yielding,
\begin{equation} \label{eq:pseudo-iso-df}
    f_z(J_z) \propto e^{-\Omega_z J_z/\sigma_z^2},
\end{equation}
which is referred to as the \textit{pseudo-isothermal} DF, where $f_z$ must be appropriately normalised so that $\int f_z \,\text{d}J_z = \frac{1}{2\pi}$.
\par
While equation \eqref{eq:pseudo-iso-df} is a widely used building block, it is not guaranteed to describe the vertical DF of a coeval population shaped by a vertical heating mechanisms. Indeed, \citet{Binney_2010} already discussed empirical modifications to the pseudo-isothermal vertical DF, showing that replacing the exponential dependence on $J_z$ with a more flexible algebraic form can improve fits to vertical density and velocity distributions inferred from the \textit{Geneva Copenhagen Survey} (GCS), particularly for small $|v_z|$. However, such modifications are primarily empirical. They are useful for fitting the present-day aggregate disc, but they do not explain what DF should arise for a coeval population under a specified scattering mechanism. This distinction is important because the Galactic disc is not a single isothermal population. It is a superposition of stellar cohorts with different ages, birth conditions, and heating histories. A present-day aggregate DF may therefore look approximately pseudo-isothermal even if the DF of each individual cohort is not. In this paper, our focus is on the basic building blocks, namely, the vertical DF for a single-age population in the presence of GMC scattering.
\par
In the next section, we provide a simple toy model, demanding the modification of the pseudo-isothermal DF in a limiting case of a razor thin GMC distribution, and then extend this to a more general configuration with a GMC distribution in height. 

\section{A new DF with GMC scattering} \label{sect: DF-GMC}
\par
We present a simple statistical argument that shows a deviation from the pseudo-isothermal DF in the simplest possible system. Consider scattering in one dimension (i.e. ignoring radial and azimuthal variations) by a razor-thin GMC population at $z=0$. For a vertical potential, $\Phi(z)$, the Hamiltonian reads,
\begin{equation} \label{eq: hamiltonian}
    H(z,v_z) = \frac12 v_z^2 + \Phi(z),
\end{equation}
with vertical action,
\begin{align}
    J_z(E_z) = \frac{1}{2\pi} \oint v_z \, dz = \frac2\pi \int_0^{z_\mathrm{max}(E_z)} \sqrt{2(E_z - \Phi(z))} \ dz,
\end{align}
where $z_\mathrm{max}$ is the turning point in $z$.
\par
Let $P(J_z,t)$ denote the cumulative distribution of actions $J_z$, namely, the fraction of stars with $J_z' \leq J_z$ at a given time, $t$. The corresponding probability density in action and energy given by $p_{J}(J_z,t) \equiv dP / dJ_z$ and $p_E(E_z,t)\equiv dP / dE_z$, respectively. The phase space density is then given by,
\begin{align} \label{eq:fz-def}
    f_z \equiv \frac{dP}{dz \ dv_z} = \frac{1}{2\pi} \frac{dP}{dJ_z} = \frac{\Omega_z(E_z)}{2\pi} \frac{dP}{dE_z},
\end{align}
where $\Omega_z(E_z) = dE_z / dJ_z$ denotes the vertical frequency, and for a phase-mixed 1D system, $f_z$ is independent of $\theta_z$.

\subsection{A generalised DF for razor-thin GMC distributions} \label{sect: razor-thin-gmc}
\par
We proceed to model the time evolution of the distribution in the presence of GMC scattering. We assume the kick in stellar vertical velocity from a GMC scattering is constant, and the star will only encounter one cloud, or equivalently, one deflection, per midplane-crossing. In a general potential, the midplane-crossing rate is a function of energy, $E_z$, with two crossings per orbital period,
\begin{align}
    \lambda(E_z) = \frac{\Omega_z(E_z)}{\pi}.
\end{align}
We formalise the midplane kick in velocity as follows, $v_{\mathrm{mid}} \mapsto v_{\mathrm{mid}} + \Delta{v_z}$. For simplicity, we first assume a normal distribution $\Delta{v_z} \sim \mathcal{N}(0,\sigma_k)$, namely, $\langle \Delta v_z \rangle = 0$, and $\langle (\Delta v_z)^2 \rangle = \sigma_k^2$ (for a more realistic configuration, \textit{see} \S \ref{sect:lacey-scattering}). In the midplane, the corresponding (vertical) energy change reads,
\begin{align}
    \Delta E_z  = v_z \, \Delta v_z + \frac{(\Delta v_z)^2}{2}.
\end{align}
\par
To leading order, the moments are, $\langle \Delta E_z \rangle = \sigma_k^2/2$, and $\langle (\Delta E_z)^2 \rangle = 2E_z \sigma_k^2$. For sufficiently small kicks, the diffusion in energy can be modelled as a stochastic differential equation (SDE) in the It\^o sense, 
\begin{align} \label{eq:energy-sde}
    dE_z = D^{(1)}(E_z) \, dt + \sqrt{D^{(2)}(E_z)} \, dW_t,
\end{align}
and,
\begin{align}
    D^{(1)}(E_z) &= \lambda(E_z)\frac{\sigma_k^2}{2} \label{eq:de-1} \\
    D^{(2)}(E_z) &= \lambda(E_z) \, 2E_z\sigma_k^2, \label{eq:de2}
\end{align}
where we multiply the energy moments by the crossing rate, $\lambda(E_z)$, to obtain the first and second order diffusion coefficients, $D^{(1)}(E_z)$ and $D^{(2)}(E_z)$, respectively. $W_t$ denotes a standard one-dimensional Brownian motion (Wiener process), so that $dW_t$ is the increment with mean $\mathbb{E}[dW_t]=0$ and variance $\mathbb{E}[(dW_t)^2]=dt$. Equation (\ref{eq:energy-sde}) is equivalent to the time evolution of the probability density in energy $p_E (E_z,t) \equiv dP/dE_z$ under the one-dimensional Fokker-Planck equation,
\begin{align}
    \frac{\partial p_E}{\partial t} = -\frac{\partial}{\partial E_z} \left[D^{(1)} p_E \right] + \frac{1}{2} \frac{\partial^2}{\partial E_z^2} \left[D^{(2)} p_E \right].
\end{align}
The solution to this differential equation, $p_E$, is then substituted into equation (\ref{eq:fz-def}) to obtain the phase-space density, or equivalently, the energy and action DFs. 
\par
Fortunately, for some potentials, as we will demonstrate below, the SDE in equation (\ref{eq:energy-sde}), can be transformed to a squared-Bessel process, which are solutions to SDEs of the form,
\begin{align} \label{eq:sde-bessel-main}
    dX_t = \delta \, dt + 2 \sqrt{X_t} \, dW_t, \quad X_0 = x,
\end{align}
where $\delta \geq0$ is a constant, and $X_0=x$ is the initial condition.
In particular, if our SDE evolution in equation \eqref{eq:energy-sde} can be related by some linear mapping
$E_z=\alpha_E X_t$ for some constant $\alpha_E>0$, and $X_t$ solves the SDE in equation \eqref{eq:sde-bessel-main}, we can transform variables in equation \eqref{eq:energy-sde}, using $dE_z = \alpha_E \, dX_t$, and $\sqrt{X_t} = \sqrt{E_z/\alpha_E}$, so that,
\begin{align} \label{eq:energy-sde-transformed}
    dE_z = \alpha_E \, \delta \, dt + 2 \sqrt{\alpha_E} \sqrt{E_z}  \, dW_t.
\end{align}
This SDE is then used to determine $\alpha_E$ by comparing to the original SDE. Finally, as derived in Section S1 of the Supplementary Material, the probability density in energy can be written as, 
\begin{align}\label{eq:gamma-e}
  p_E(E_z,t)=\frac{1}{\Gamma(\delta/2)\,(2\alpha_E t)^{\delta/2}}
  \,E_z^{\delta/2-1}\exp\!\left(-\frac{E_z}{2\alpha_E t}\right),
\end{align}
for $E_z>0$, and $\alpha_E$ determined. $\Gamma$ is the Gamma-function.
\par
Now, we proceed with a derivation for the action DFs in the two analytic cases for vertical potentials, which bracket more realistic discs: the harmonic oscillator potential, and the razor-thin (linear) potential.

\subsubsection{Harmonic oscillator}
\par
Generally, a near-isothermal system will approach near-constant mass density sufficiently close to the midplane, and thus approximately a harmonic potential,
\begin{align}
    \Phi(z) = \frac12\Omega_z^2 z^2.
\end{align}
\par
Here, the midplane-crossing rate $\lambda = \Omega_z / \pi$ is a constant, so the SDE in equation \eqref{eq:energy-sde}, using the definitions of moments in eq. \eqref{eq:de-1}--\eqref{eq:de2},
reduces to,
\begin{align}
    dE_z = \frac{\Omega_z \sigma_k^2}{2\pi} dt + \sqrt{\frac{2\Omega_z\sigma_k^2}{\pi} E_z} \, dW_t.
\end{align}

Comparison with equation \eqref{eq:energy-sde-transformed} yields $\alpha_E = \frac{\Omega_z \sigma_k^2}{2\pi}$ and $\delta = 1$. Substitution into equation \eqref{eq:gamma-e} gives,
\begin{align}
    p_E(E_z,t)=\frac{1}{\sqrt{\Omega_z  \sigma_k^2  t}}
  \,E_z^{-1/2}\exp\!\left(-\frac{\pi E_z}{\Omega_z \sigma_k^2 t}\right).
\end{align}
Finally, we substitute into equation \eqref{eq:fz-def} to obtain the energy DF,
\begin{align}
    f_z(E_z,t)=\frac{\Omega_z^{1/2}}{2\pi\sqrt{\sigma_k^2  t}}
  \,E_z^{-1/2}\exp\!\left(-\frac{\pi E_z}{\Omega_z \sigma_k^2 t}\right)
\end{align}
If we define an effective velocity dispersion,
\begin{align}
    \sigma_z^2(t) \equiv \frac{\Omega_z \sigma_k^2}{\pi} t,
\end{align}
then the energy DF can be written as,
\begin{align} \label{eq:modified-df-Ez-HO}
    f_z(E_z) = \frac{\Omega_z}{2\pi^{3/2}\sigma_z} E_z^{-1/2} e^{-E_z/\sigma_z^2},
\end{align}
and the vertical action DF follows immediately via $J_z = E_z/\Omega_z$,
\begin{align} \label{eq:modified-df-Jz-HO}
    f_z(J_z) = \frac{\Omega_z^{1/2}}{2\pi^{3/2}\sigma_z} J_z^{-1/2} e^{-\Omega_zJ_z/\sigma_z^2},
\end{align}
This extra factor of $J_z^{-1/2}$ implies a steeper cusp at smaller actions, than the pseudo-isothermal DF. Finally, the velocity dispersion grows like $\sigma_z \propto \sqrt{t}$, which is expected for a random walk in velocity. 

\subsubsection{Razor-thin potential}
\par
If by contrast all the mass is concentrated in a razor-thin midplane, we have,
\begin{align}
    \Phi(z)=K|z|.
\end{align}
Here, the vertical frequency, and thus midplane crossing rate, is a decreasing function of energy,
\begin{align}
\Omega_z(E_z)=\frac{\pi K}{2\sqrt{2}} E_z^{-1/2} \,\,\, \mathrm{,} \quad  \lambda(E_z) = \frac{K}{2\sqrt{2}} E_z^{-1/2}.
\end{align}
As before, we substitute $\lambda(E_z)$ into equations \eqref{eq:de-1}--\eqref{eq:de2} to obtain the energy SDE
\begin{align}\label{eq:sde-linear-E}
    dE_z
    =\frac{K\sigma_k^2}{4\sqrt{2}}\,E_z^{-1/2}\,dt
    +\sqrt{\frac{K\sigma_k^2}{\sqrt{2}}}\,E_z^{1/4}\,dW_t.
\end{align}
While an immediate comparison to equation \eqref{eq:sde-bessel-main} does yield $\alpha_E$, we can transform variables:
$Y_t \equiv E_z^{3/2}$, so that $dY_t = \tfrac{3}{2} E_z^{1/2} dE_z + \tfrac38 E_z^{-1/2}D^{(2)}(E_z) \, dt$. This yields an SDE in $Y_t$,
\begin{align}\label{eq:sde-linear-Y}
    dY_t
    =\frac{3K\sigma_k^2}{4\sqrt{2}}\,dt
    +\frac{3}{2}\sqrt{\frac{K\sigma_k^2}{\sqrt{2}}}\,\sqrt{Y_t}\,dW_t.
\end{align}
Now define a rescaled variable $X_t=\alpha_Y Y_t$, and with a choice, $\alpha_Y \equiv \frac{16\sqrt{2}}{9K\sigma_k^2}$, equation \eqref{eq:sde-linear-Y} becomes the squared-Bessel SDE,
\begin{align}
    dX_t = \delta\,dt + 2\sqrt{X_t}\,dW_t,
\end{align}
with $\delta=\frac{4}{3}$. 
\par
If $X_t$ is a squared-Bessel process, then a linear rescaling of $X_t$ has the Gamma-form density given in equation~(S12) of the Supplementary Material. Here, $Y_t=X_t/\alpha_Y$, so equation (S12) applies with scale parameter $2t/\alpha_Y$. Transforming back to energy using $Y_t=E_z^{3/2}$ gives,
\begin{align}\label{eq:pe-linear}
    p_E(E_z,t)
    =\frac{3}{2\,\Gamma(2/3)\,E_0(t)}\,
    \exp\!\left[-\left(\frac{E_z}{E_0(t)}\right)^{3/2}\right],
\end{align}
for $E_z > 0$. The characteristic energy scale is given by,
\begin{align}\label{eq:E0-linear}
    E_0(t)\equiv \left(\frac{2t}{\alpha_Y}\right)^{2/3}
    =\left(\frac{9K\sigma_k^2}{8\sqrt{2}}\,t\right)^{2/3}.
\end{align}
The corresponding energy DF follows from equation \eqref{eq:fz-def},
\begin{align}\label{eq:fE-linear}
    f_z(E_z,t)=\frac{3K}{8\sqrt{2}\,\Gamma(2/3)\,E_0(t)}\, E_z^{-1/2}
    \exp\!\left[-\left(\frac{E_z}{E_0(t)}\right)^{3/2}\right].
\end{align}
\par
Finally, to obtain the action DF, we use the analytic action--energy relation for the linear potential,
\begin{align}
    J_z(E_z)=\frac{4\sqrt{2}}{3\pi K}\,E_z^{3/2}.
\end{align}
Since $E_z^{3/2}$ is proportional to $X_t$, it follows that $J_z$ is \emph{linearly} related to $X_t$, $J_z=\alpha_J X_t$, where $\alpha_J=\frac{3\sigma_k^2}{4\pi}$ and $\delta = 4/3$, so that the Gamma shape parameter is $\delta/2 = 2/3$. Hence, equation (S12) can be applied directly to $J_z$, yielding,
\begin{align}\label{eq:pJ-linear}
    p_J(J_z,t)
    =\frac{1}{\Gamma(2/3)\,J_0(t)^{2/3}}
    J_z^{-1/3}\exp\!\left(-\frac{J_z}{J_0(t)}\right)
\end{align}
for $J_z > 0$, with characteristic action scale,
\begin{align}
    J_0(t)\equiv 2\alpha_J t=\frac{3\sigma_k^2}{2\pi}\,t.
\end{align}
Finally, for a phase-mixed 1D system the action DF is given by equation \eqref{eq:fz-def}, 
\begin{align}\label{eq:fJ-RZ}
    f_z(J_z,t)=\frac{1}{2\pi \,\Gamma(2/3)\,J_0(t)^{2/3}}
    J_z^{-1/3}\exp\!\left(-\frac{J_z}{J_0(t)}\right).
\end{align}
Because of the different dependence $J(E_z)$ the cusp in actions is slightly weaker ($J_z^{-1/3}$) than in the harmonic potential.

Finally, the characteristic action scale for both potentials grows as $\propto t$, while the energy scale in a linear potential grows more slowly, $E_0(t) \propto t^{2/3}$. The reason the energy and action distributions behave differently is that the scattering is applied at the midplane, so the kick statistics are expressed in terms of the midplane velocity and the crossing rate. In the harmonic potential, the crossing rate is constant, so the energy itself is the diffusing variable and one obtains an exponential tail in $E_z$ with a scale $\propto t$. Conversely, for the razor-thin potential, the crossing rate increases at low energies, $\lambda(E_z)\propto E_z^{-1/2}$, and the SDE closes instead for the combination $Y\equiv E_z^{3/2}\propto J_z$. Thus, the action (or equivalently $E_z^{3/2}$) behaves like a squared-Bessel diffusion and retains a modified exponential in $J_z$, with a cusp $p_J\propto J_z^{\delta/2-1}$ at small actions. The energy PDF is then obtained by a nonlinear change of variables $J_z\propto E_z^{3/2}$, which can remove the cusp in $p_E$. In fact, $p_E$ becomes cored for the razor-thin case, even though the underlying action distribution remains cuspy. 
\par
While we have only investigated the action DFs for two potentials, we suspect the action DFs will be cuspier than the pseudo-isothermal DF at small actions. In general, the small-$J_z$ behaviour is controlled by the scaling of the crossing rate $\lambda(E_z)\propto \Omega_z(E_z)$ and by the action--energy relation $J_z(E_z)$. If these enter the SDE such that some power of $E_z$ evolves as a square-root (Bessel) diffusion, then the resulting $p_J$ will contain the factor, $p_J\propto J_z^{\delta/2-1}$ and hence $f_z\propto J_z^{\delta/2-1}$. For $\delta < 2$, the DF will exhibit a cusp. However, if the kick amplitude or kick rate depends strongly on phase or velocity, or if additional physics regularises the dynamics near $J_z=0$ (e.g.\ finite GMC scale height or non-instantaneous interactions), the cusp can be weakened or removed. We will investigate scattering in an extended GMC distribution in \S \ref{sect:lacey-scattering}, and find in more realistic configurations that this cusp is still present.

\subsection{Test-particle simulation}
To verify the isothermal DF is not a suitable fit to vertical action distributions in the presence of GMC scattering, we perform a one-dimensional test-particle simulation with $N=10^6$ stars, which are left to evolve in phase space under the vertical frequency, $\Omega_z = 71$ km s$^{-1}$ kpc$^{-1}$. in a harmonic oscillator potential, and the only source of diffusion is from GMC scattering in the Galactic midplane. We adopt code units, $z_0 = 0.3$ kpc, and $V_0 = 30$ km s$^{-1}$, so that one time unit is $T_0 = z_0/V_0 = 9.778$ Myr. In these units, $\Omega_z = 0.71$. We run the simulation for a total time of $t=1000$, which corresponds to 9.78 Gyr. We additionally perform a test-particle simulation in the razor-thin potential, with $K=0.5$, which corresponds to $K_{\rm phys} = 1500$ km$^{2}$ s$^{-2}$ kpc$^{-1}$, in physical units. We assume the star-cloud interaction takes a Gaussian random walk in velocity, ignoring scattering in $z$, as described in \S \ref{sect: razor-thin-gmc}.

\begin{figure}
    \centering
    \includegraphics[width=\linewidth]{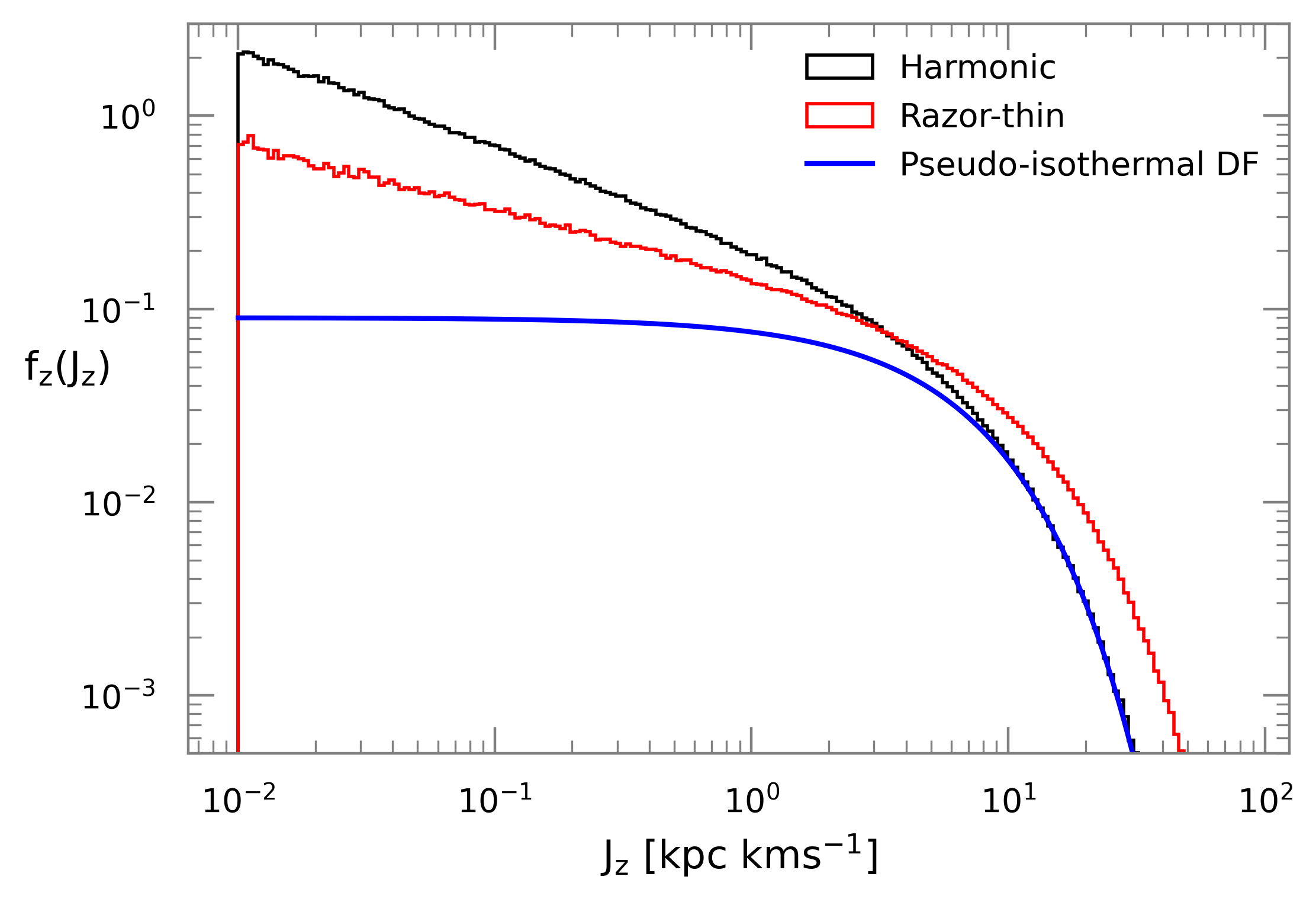}
    \caption{A histogram of the distribution function, $f_z$, at $t\approx 10$ Gyr, from a one-dimensional test-particle simulation with $N=10^6$ particles, with a razor-thin GMC distribution in the Galactic midplane, in a harmonic oscillator potential (black), and the linear potential (red). The pseudo-isothermal (blue) DF from equation \eqref{eq:pseudo-iso-df} is also plotted for reference.}
    \label{fig:DF_fit}
\end{figure}

\par
\begin{figure}
    \centering
    \includegraphics[width=\linewidth]{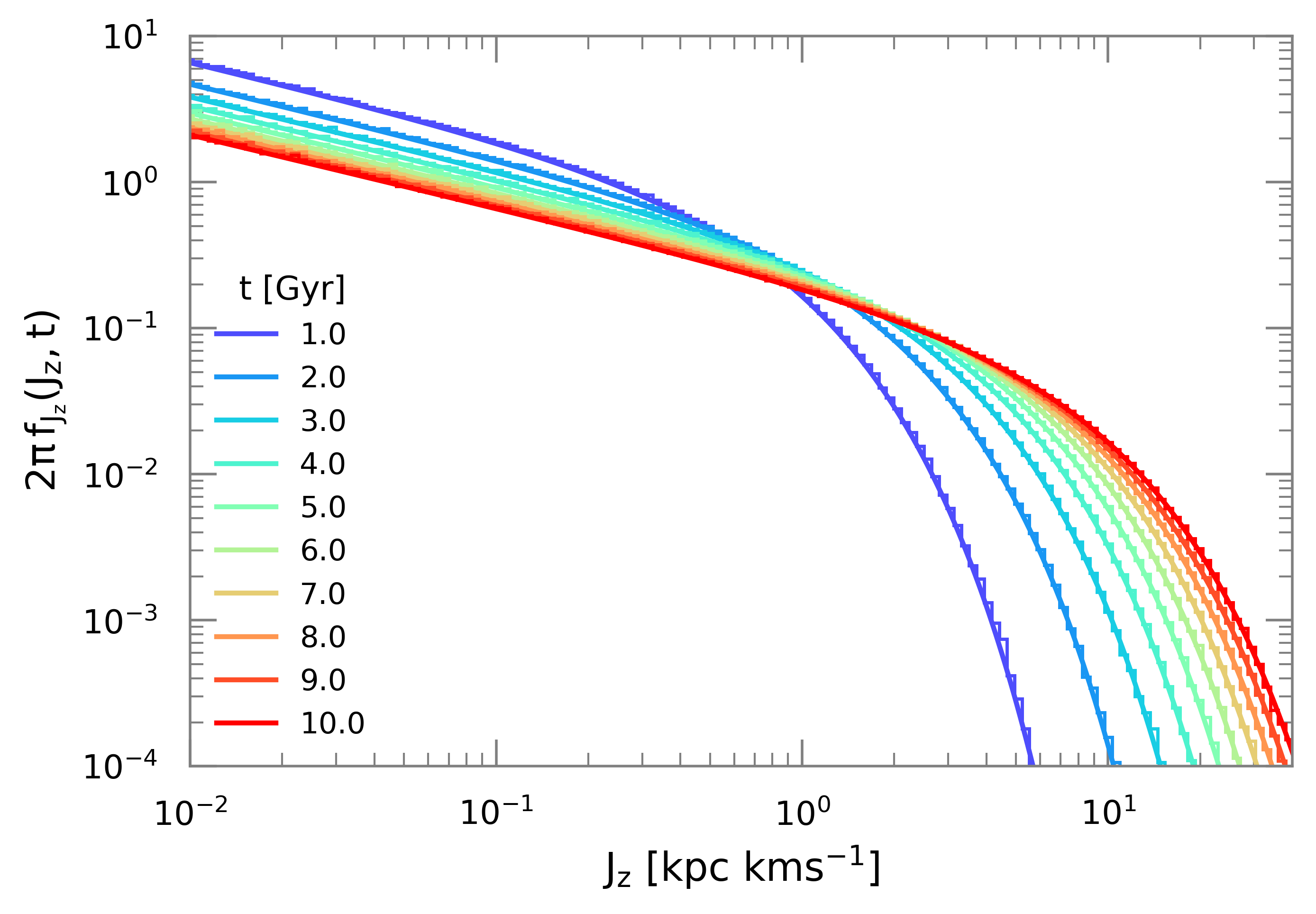}
    \caption{Time evolution of the vertical action distribution, $J_z$, in a harmonic oscillator potential, with the exact theoretical curves of equation \eqref{eq:modified-df-Jz-HO} plotted as solid lines.}
    \label{fig:ho-time-ev}
\end{figure}
\begin{figure}
    \centering
    \includegraphics[width=\linewidth]{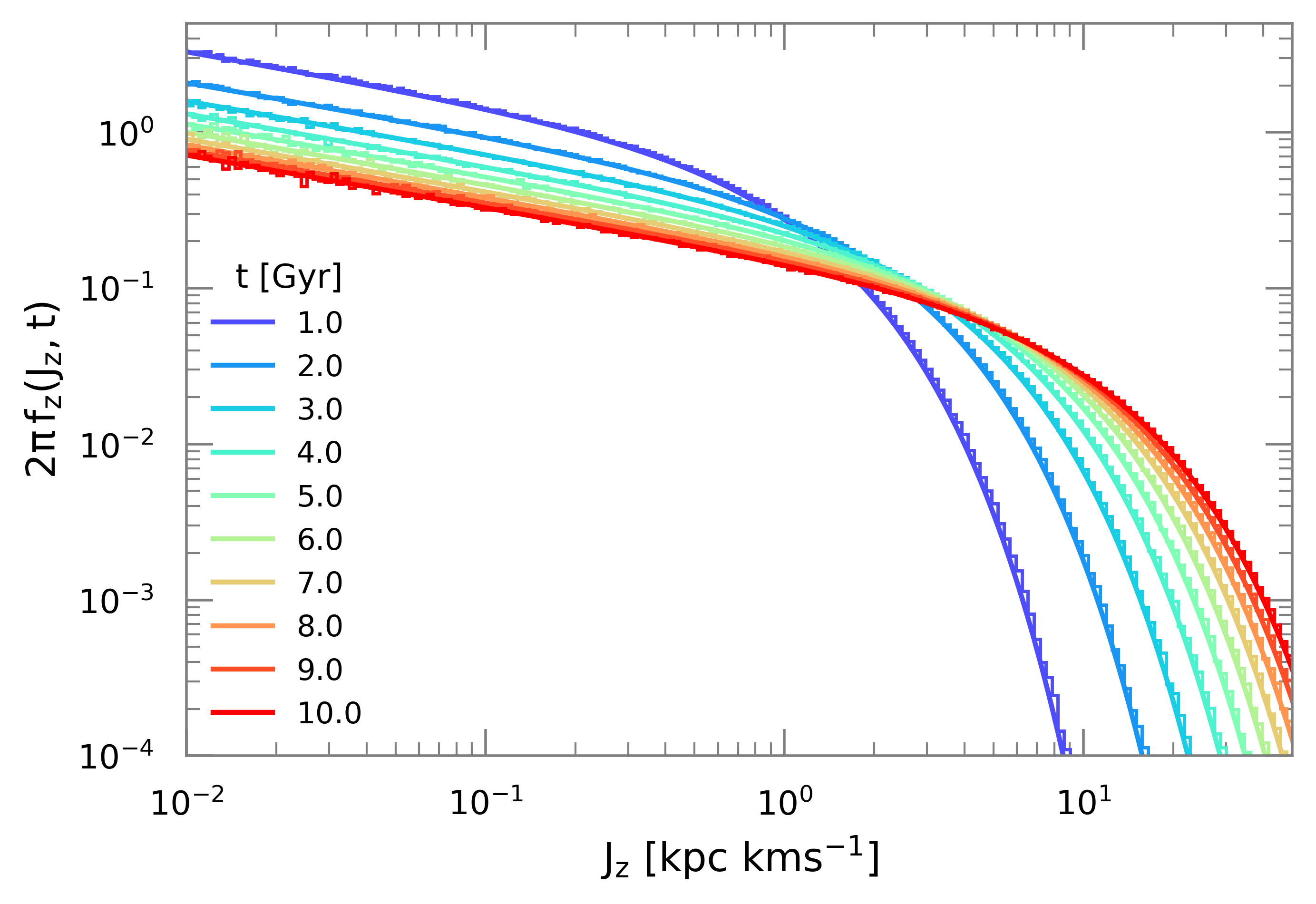}
    \caption{Same as Figure \ref{fig:ho-time-ev}, but for the razor thin potential, and theoretical curves of equation \eqref{eq:fJ-RZ} plotted as solid lines.}
    \label{fig:rz-time-ev}
\end{figure}

We initialise particles with a cold distribution ($J_z=0$), but add a tiny jitter in action ($\sim 10^{-4}$) for numerical stability, and sample angles from a uniform distribution, $\theta_z \in [0,2\pi]$. In the simulation, each particle experiences one deflection in velocity only, per midplane crossing, and we set the dispersion of the kick to $\sigma_k = 0.05$ in code units. Figure \ref{fig:DF_fit} displays a histogram of the vertical action DF at $t=10$ Gyr, for the simulations in both potentials. For reference, the pseudo-isothermal DF is plotted in blue, and both simulations demonstrate a steeper distribution at small actions. Moreover, in Figure \ref{fig:ho-time-ev}, we plot the time evolution of the vertical action DF in the harmonic oscillator potential, and plot in solid lines the theoretical DFs derived in equation \eqref{eq:modified-df-Jz-HO}. Figure \ref{fig:rz-time-ev} displays the same time evolution, but in the razor-thin potential, and with theoretical curves of equation \eqref{eq:fJ-RZ} plotted with solid lines. Our test-particle simulations therefore demonstrate an agreement with the predicted time-evolution of the vertical action distribution.

Crucially, we emphasise the the solutions derived in this section are \textit{not} steady state DFs, but rather time-dependent solutions. Despite the kicks being unbiased in velocity, $\langle v_z\rangle =0$, there is an induced bias in energy, such that, $\langle E_z\rangle = \sigma_k^2/2 > 0$. The razor-thin GMC configuration therefore injects vertical energy indefinitely, and the characteristic energy and action scales continue to grow with time, rather than reaching an asymptotic value. This absence of a stabilising drift in $E_z$ and $J_z$ is precisely what motivates a more realistic treatment in \S \ref{sect:lacey-scattering}, where diffusion coefficients are computed for an extended GMC distribution, and acquire non-linear velocity dependence. In that case, we show the system can settle to a steady state, while still retaining signatures of non-isothermality at low $J_z$.

\section{Velocity dependent GMC scattering} \label{sect:lacey-scattering}
\par
The analytical form of the DF in \S \ref{sect: DF-GMC} is only exact in the idealised limit where the scattering probability is independent of velocity, and occurs exactly once per midplane-crossing at $z=0$. Here, we extend the simple simulation in \S \ref{sect: DF-GMC}, by allowing the scattering probability to depend on the star's phase-space coordinates. This begins with implementing a GMC distribution in height, and determining velocity moments from the mechanics of two-body scattering.

\subsection{Weak encounters}
\par
To model the star-cloud interactions from weak encounters alone, we simulate the master equation for the phase-space DF $f$ \citep[see][\S7.4]{BT2008}
\begin{align} \label{eq:master}
    \frac{\partial f}{\partial t}=\Gamma[f],
\end{align}
where,
\begin{align} \label{eq:encounter-operator}
        \Gamma[f]&\equiv
    \int\mathrm{d}^2(\Delta{\mathbf{w}})\,\big[ \psi(\mathbf{w}-\Delta{\mathbf{w}}, \Delta{\mathbf{w}})\,f(\mathbf{w}-\Delta{\mathbf{w}}) - \psi(\mathbf{w},\Delta{\mathbf{w}})\,f(\mathbf{w})\big]
\end{align}
where $\mathbf{w}=(z,v_z)$, and $\psi(\mathbf{w},\mathbf{\Delta w})\rm d^2 (\mathbf{\Delta w}) \Delta t$ is the probability a star at phase-space coordinate $\mathbf{w}$ is scattered to the phase space volume $d^2 (\mathbf{\Delta w})$ centred around $\mathbf{w+\Delta w}$ in a time interval $\Delta t$. We remark, the orbital evolution terms are treated separately and the $\Gamma[f]$ term accounts only for stochastic evolution. We neglect scattering in $z$, and assume the star experiences a deflection purely in velocity, while the recoil of the cloud itself is negligible. Hence $\Delta \mathbf{w} = (0,\Delta{v_z})$, and $\psi(\mathbf{w},\mathbf{\Delta w}) = \delta(\Delta z) \, \psi(\Delta v_z | z,v_z)$, so that the DF evolution then simplifies to,
\begin{align} \label{eq:master-1d}
    \frac{\mathrm{d}f}{\mathrm{d}t} = \int \mathrm{d}(\Delta{{v}})\,\big[ \psi(\Delta {v} | z, v- \Delta{{v}})\,f({v}-\Delta{{v}}) - \psi(\Delta{v}| z,v)\,f({v})\big],
\end{align}
where we dropped subscripts on $v_z$, since the scattering is one-dimensional.
If we define, the $n-$th Kramers-Moyal moment \citep{Spitzer_1987,BT2008} as, 
\begin{align} 
    D^{(n)}_{v^n}(z,v_z) \equiv \int (\Delta v_z)^n \, \psi(\Delta{v}| z,v_z) \, \text{d}(\Delta v_z),
\end{align}
then the drift and diffusion coefficients are given, respectively as, $D^{(1)}_v(z,v_z)$ and $D^{(2)}_{vv}(z,v_z)$. 
For weak encounters, we assume  $|\Delta v_z|$ is small, and in particular, $|\Delta v_z| \ll V_{\rm{rel}}$, where $V_{\rm{rel}}$ is the magnitude of relative velocity between a star and cloud. Expanding equation (\ref{eq:master-1d}) as a Taylor series, neglecting terms of $\mathcal{O}(\Delta v^3)$, yields the Fokker-Planck equation in one dimension for the velocity,
\begin{align}
    \frac{\partial f}{\partial t}=-\frac{\partial}{\partial v_z}\left[D^{(1)}_v(z,v_z) f \right]
    +
    \frac{1}{2}\frac{\partial^2}{\partial v_z^2}
    \left[D^{(2)}_{vv}(z,v_z) f \right].
    \label{eq:FP-velocity}
\end{align}
in terms of the drift and diffusion coefficients. Equation (\ref{eq:FP-velocity}) describes the stochastic evolution due to star-cloud interactions. Deterministic phase-space evolution, namely, the terms associated with the smooth vertical potential, is treated separately and is not included in  equation (\ref{eq:FP-velocity}).
\par
To model weak encounters, we adopt a Monte Carlo approach to evolve the DF. For a sufficiently large number of particles, $N$, and a reasonable timestep, $\Delta t$, we can model the Fokker-Planck equation stochastically. We must choose $\Delta t$ that is short compared to the orbital timescale, but sufficiently long so that the cumulative effect of many unresolved weak encounters over $\Delta t$ is well approximated as a Gaussian. Namely, the probability of velocity changes, $\Delta v$, is Gaussian, with a mean, $\mu = \langle \Delta v_z \rangle$ and variance, $\sigma^2 = \langle \Delta v_z^2\rangle- \langle \Delta v_z\rangle^2$. These are related directly to the drift and diffusion coefficients as $\langle \Delta v_z \rangle = D^{(1)}_v\Delta t$ and $\langle \Delta v_z^2 \rangle = D^{(2)}_{vv}\Delta t$. Hence, stars are integrated in a given potential, and at each time interval, $\Delta t$ the velocity deflection is sampled from a normal distribution, centred at $\mu$ with variance, $\sigma^2$. This stochastic evolution is equivalent to a (discrete) SDE in the It\^{o} sense, to leading order, $\Delta t$,
\begin{align}
    \Delta z &= v_z \, \Delta t  \label{eq:sde1}\\
    \Delta v_z &= -\partial_z \Phi(z) \, \Delta t + D^{(1)}_v(v) \,  \Delta t + \sqrt{D^{(2)}_{vv}(v)} \label{eq:sde2} \,  \Delta W_t,
\end{align}
where $ \Delta W_t$ represents increments of a \textit{Wiener process}, satisfying $\mathbb{E}\{ \Delta W_t\} = 0$ and $\mathbb{E}\{( \Delta W_t)^2\} =  \Delta t$.  The details of the test-particle simulation are given in \S \ref{sect:new-test-particle-sim}, and we compute the moments in the next subsection. 

\subsubsection{Moments of velocity change} \label{subsect:moments of velocity}
\par
At each instance of a velocity deflection, the magnitude of the velocity scatter is sampled from a normal distribution, where the first and second moments are to be determined. Rather than assigning a constant value, as in \S \ref{sect:separable-df}, we will more accurately derive expressions based on simple mechanics of two-body encounters (\textit{see} \citep{Henon_1973,Lacey_1984,Ida_1993} for analogous derivations). We assume all GMCs have the same mass, $M_c$, with a number density $n_c$, and all test stars have the same mass $m_1$, with $M_c \gg m_1$. 
\par
Denote $\mathbf{V}_c = V_c \,\hat{\mathbf n}$, as the 3D cloud velocity, where $V_c = |\mathbf{V}_c|$ is the cloud speed and $\hat{\mathbf n}$ is a unit vector specifying its direction. Assuming an isotropic velocity distribution, the velocity magnitude $V_c$ is sampled from a Maxwellian \citep{Gieles_2006}, whose distribution is,
\begin{equation} \label{eq:cloud-maxwellian}
    f({V}_c) \, \text{d}{V}_c = \frac{4\pi V_c^2}{(2\pi \sigma^2)^{3/2}} \exp \left(-\frac{V_c^2}{2\sigma^2} \right) \, \text{d}{V}_c ,
\end{equation}
where $\sigma$ is the 1D cloud velocity dispersion. Since our simulation is one-dimensional, we draw in-plane components from some distribution $\mathbf{g(v}_{\perp}| \tau)$, where $\mathbf{v}_{\perp} = (v_x,v_y)$ are in-plane velocities of the star. The in-plane velocities will also implicitly be a function of age, $\tau$ as stellar populations heat over time. The ensemble relative speed is defined as,
\begin{align} \label{eq:v_rel_integral}
    \langle V_{\rm rel}(v_z | \tau)\rangle =\int d^2\mathbf v_\perp\,& g(\mathbf v_\perp|\tau)
\int_0^\infty dV_c \, f(V_c) \nonumber \\
 \frac{1}{4\pi} \int  &d\Omega \,
\left|V_c\hat{\mathbf n}-\mathbf v_\star\right|,
\end{align}
with $d\Omega$ is the solid-angle element, $\mathbf{v_\star}=(v_x,v_y,v_z)$ is the stellar velocity, and both velocities are computed in the local standard of rest (LSR). We assume the GMCs do not heat significantly over time \citep{Heyer_Dame_2015}, so that $\sigma$ remains constant. We choose $\sigma = 6$ km s$^{-1}$ \citep{Dobbs_Pringle_2011,Heyer_Dame_2015}. The closed form solution of the integral in equation (\ref{eq:v_rel_integral}) is given in Appendix \ref{app: relative speed}. In practice, for simplicity and computational efficiency, we adopt the RMS approximation
\begin{align} \label{eq:v_rms}
    \langle V_{\rm rel}(v_z|\tau) \rangle \approx V_{\rm rms} = C_V(v_z,\tau) \sqrt{v_z^2 + 3\,\sigma^2 + \sigma_R^2(\tau) + \sigma_{\phi}^2(\tau)},
\end{align}
where $\sigma_R$ and $\sigma_{\phi}$ are stellar in-plane velocity dispersions, which increase over time. $C_V(v_z,\tau)$ is an additional (dimensionless) correction factor given in Appendix \ref{app: relative speed}. We also justify the use of this approximation in the appendix, and explain how in-plane velocities are drawn.
\par 
The first and second-order moments of vertical velocity are derived analytically in Appendix \ref{app:moments-derivation}, and their exact expressions are given in equations (\ref{eq: dv-exact}) and (\ref{eq:dv_sq-exact}), respectively. In practice, we adopt useful approximations, justified in Appendix \ref{app:moments-derivation},
\begin{equation} \label{eq:dv}
    \langle \Delta v_z \rangle \approx -2\pi G^2 M_c^2 n_c \Delta t  \ln \left [ 1 +\Lambda ^2 \right] \frac{v_z}{V_{\rm rms}^3}
\end{equation}
\begin{align} \label{eq:dv_sq}
    \langle (\Delta v_z )^2 \rangle \approx & \frac{4\pi G^2 M_c^2 n_c \Delta t}{V_{\rm rms}} \times \nonumber \\
    &\left[ \frac{v_z^2}{V_{\rm rms}^2 }\frac{\Lambda^2}{1+\Lambda^2} + \frac12 \left( 1 -  \frac{v_z^2}{V_{\rm rms}^2} \right  ) \left[ \ln \left ( 1 +\Lambda^2 \right) - \frac{\Lambda^2}{1+\Lambda^2}  \right] 
    \right],
\end{align}
where,
\begin{equation} \label{eq:coulomb-log}
    \Lambda \equiv \frac{b_{\mathrm{max}}}{b_{\mathrm{min}}},
\end{equation}
and $\ln \Lambda$ is referred to as the Coulomb logarithm; the log of the ratio of maximum to minimum impact parameters, and $V_{\rm rms}$ is given by equation (\ref{eq:v_rms}). These approximations are consistent with the results of \citet{Lacey_1984} and \citet{Ida_1993}, yet the approximation is prone to errors, particularly in the limit of small vertical velocities. This is explained in Appendix \ref{app:moments-derivation}, and corrections to these approximations are given. Equations (\ref{eq:dv}) and (\ref{eq:dv_sq}) are derived assuming the \textit{local approximation} holds, that is, the mean separation of GMCs is much smaller than the scale length on which the GMC density varies. We also assume the \textit{impulse approximation}, where scattering in $z$ is negligible and the star experiences a deflection in velocity only, which is valid if the encounter time is much shorter than the crossing time. 
\par
Unfortunately, the Coulomb logarithm, and in particular, the choice of maximum and minimum impact parameters, is still not adequately defined in the literature, and is rather arbitrarily chosen (e.g. \citet{Penarrubia_2019}, and references therein). The dominant logarithmic dependence in the moments enters through terms of the form $\ln(1+\Lambda^2)$, which approach $2\ln\Lambda$ for $\Lambda\gg1$ and diverge for small and large impact parameters \citep{Chandrasekhar_1943}. Thus we cannot in theory integrate $b$ in the range $[0, \infty]$. Typically, $b_{\mathrm{min}}$ is taken as the distance where there is a maximum deflection of $90^\circ$, and $b_{\mathrm{max}}$ is some typical size of the system. Recently, \citet{Penarrubia_2019} argued the Coulomb logarithm can be defined as the ratio of average distance between substructures and their sizes, namely, $\ln \Lambda  = \ln (D/c)$, where $D$ is the average separation of substructures and $c$ is their characteristic size. In the context of star-cloud scattering, there has been little discussion on the choice of impact parameters; \citet{Lacey_1984} took $b_{\mathrm{min}} = \max \left[r_c, b_{90}\right]$, where $r_c$ is the physical radius of GMCs (assumed roughly homogeneous), and $b_{90} =GM_c/V_{\rm rel}^2$ corresponds to the distance at which a $90^\circ$ deflection occurs. Lacey defined $b_{\mathrm{max}}$ as the size of an epicyclic orbit, namely, $b_{\mathrm{max}} \sim (2E_z)^{1/2} / \Omega_z$, where $\Omega_z$ is the vertical frequency. The trouble with the latter is that for stars with initially small vertical energies, $E_z$, $b_{\mathrm{max}} < b_{\mathrm{min}}$ and thus the Coulomb logarithm is ill-defined. In light of this, we propose the following choice of impact parameters,
\begin{align} 
    b_{\mathrm{min}} &= \max \left[r_c, GM_c/ V_{\rm rms} ^2 \right] \label{eq:bmin}\\
    b_{\mathrm{max}} &= \min \left[4h_c, \left(\frac{4\pi n_c}{3}\right)^{-1/3} \right] \label{eq:bmax}
\end{align}
where $b_{\mathrm{max}}$ is defined as the minimum of either $4\times$ the cloud scaleheight, $h_c$ or the mean cloud spacing,  $(4\pi n_c/3)^{-1/3}$. For a scale-height of $\sim 50$ pc, this corresponds to $b_{\rm max}$ parameters of order $\sim 200$ pc, which is much smaller than typical choices of $b_{\rm max}$. $h_c$ is defined in equation (\ref{eq:cloud-density}) via the cloud density, below, and we use the RMS speed defined in equation (\ref{eq:v_rms}) to estimate $b_{90}$. In this context, we argue that we must impose a small allowed range of impact parameters; for large impact parameters, the stars \textit{see} a mean field, that does not contribute to diffusion. Namely, in the large-$N$ limit, stars interact simultaneously with many clouds, and in this regime, the ensemble-averaged force acts like a smooth background potential, which we denote the\textit{ mean field}, and only fluctuations about this mean will contribute to heating.

With a choice of $b_{\rm max} \gg n_c^{-1/3}$, we are no longer in the case of isolated star-cloud encounters, and rather sampling from a background force from many perturbers. The perturber's gravity varies slowly so that, in this adiabatic limit, action is approximately conserved and these encounters contribute negligibly to diffusion. The question remains:\textit{ when can the mean field cause heating?} A smooth GMC background contributes to secular heating only if it varies on orbital timescales, for example, through rapid growth or decay of the GMC layer, or through transient, non-axisymmetric structure. Then the background potential does work and we can get resonant or secular heating. In our case, with a slowly time-varying vertical Gaussian GMC layer, the mean field just sets the restoring force and it does not contribute to diffusion. Thus, we justify our choice of impact parameters. Unless otherwise stated, we adopt $r_c = 50$ pc \citep{Heyer_Dame_2015}. 

\subsection{Stationary solutions}
\par
Unlike the razor-thin GMC model of \S \ref{sect: DF-GMC}, this extended GMC model does not imply monotonic energy growth. The first-order velocity diffusion coefficient, $D_v^{(1)}$ is frictional, while $D_{vv}^{(2)}$ drives diffusion, and thus the drift in energy and action is no longer monotonic. At low velocities, the diffusive term dominates, whereas at large velocities, this is offset or even outweighed by the drag term. This subtlety is crucial to allowing the flux to vanish, and thus allowing a stationary DF to exist. Stationary solutions of the Fokker-Planck equation are explored in this section.
\par
The 1D Fokker-Planck equation in velocity, from equation (\ref{eq:FP-velocity}), can be written in flux form as,
\begin{align}  \label{eq:fp-flux-form}
    \frac{\partial f}{\partial t} = -\frac{\partial}{\partial v_z} F(z,v_z,t),
\end{align}
where,
\begin{align} \label{eq:flux-term}
    F(z,v_z,t)\equiv D^{(1)}_v(z,v_z,t) f
- \frac{1}{2}\frac{\partial}{\partial v_z}
\left[D^{(2)}_{vv}(z,v_z,t)f\right],
\end{align}
is the diffusive \textit{flux}. The diffusion coefficients now generally depend on time.  We compute the stationary solution of the frozen Fokker-Planck operator, obtained by evaluating the diffusion coefficients at a \textit{fixed} age. This should not be interpreted as the final state of the fully time-dependent problem, since the coefficients depend, in general, on a varying relative velocity distribution with age, through the in-plane age--velocity relation (AVR), and the decaying GMC density with time. Instead, we derive the zero-flux equilibrium associated with fixed scattering coefficients, to investigate the phase space dependence. Henceforth, we drop all explicit time dependence, as we will evaluate the drift and diffusion coefficients, $D^{(1)}_v(z,v_z |\,t_\star)$ and $D^{(2)}_{vv}(z,v_z |\,t_\star)$, respectively, at present day $t=t_\star$,  using a present-day molecular GMC surface density of $\Sigma_{\rm GMC}\approx 5~{\rm M_\odot~pc^{-2}}$ \citep{Sharma_2021}, and assume no in-plane heating over time. Setting ${\partial f}/{\partial t} = 0$, implies a constant diffusive flux, $F$, in time. To satisfy boundary conditions, where $f\rightarrow 0$ as $|v_z|\rightarrow \infty$, the only choice of constant is $F=0$. This yields the zero-flux solution at fixed $z$,
\begin{align} \label{eq:ss-velocity}
    f(v_z \, | \, z) \propto \frac{1}{D^{(2)}_{vv}(z,v_z)} \, \exp \left( \int^{v_z} \frac{2 D^{(1)}_v(z,v')}{D^{(2)}_{vv}(z,v')} dv' \right).
\end{align}
Evidently, for approximately linear drag and a constant diffusion coefficient, we recover the isothermal velocity distribution, which is a Gaussian. However, this fixed-$z$ solution does not yet account for orbital motion in the potential, namely, the $z$-dependence, which motivates the orbit averaging below.
\par
Since changes in the DF are small over a single orbital period, and the relaxation time is several orders of magnitude larger than crossing times, it is useful to orbit average the Fokker-Planck equation. This is most easily done if we begin with the Fokker-Planck equation in action-angle coordinates,
\begin{align} \label{FP:AA}
\frac{\partial f}{\partial t}  +  \Omega_i(J)
\frac{\partial f}{\partial \theta_i} = \Gamma[f].
\end{align}
To orbit-average, we integrate over all angles, and thus the second term on the LHS vanishes by periodicity, since it is just a total derivative of a periodic function in $\theta_i$. Hence, we have orbit-averaged Fokker-Planck equation \citep[see][\S 7.4.2]{BT2008},
\begin{align} \label{eq:FP-action-orbit-average}
     \frac{\partial f}{\partial t} = - \frac{\partial}{\partial J_i} \left\{ f \overline{D^{(1)}_i}(J)\right \} + \frac12 \frac{\partial^2}{\partial J_i \partial J_j} \left\{ f \overline{D^{(2)}_{ij}}(J)\right\}
\end{align}
where the orbit-averaged diffusion coefficients are, 
\begin{align}
    \overline{D^{(1)}_i} (J)  = \frac{1}{(2\pi)^3} \int d^3 \boldsymbol{\theta} \, {D^{(1)}_i} (J,\theta) ,
\end{align}
and a similar expression for $\overline{D^{(2)}_{ij}}(J)$. 
\par
Now, in our one-dimensional case, we can derive the diffusion coefficients $\overline{D^{(1)}_J} (J_z)$ and $\overline{D^{(2)}_{JJ}} (J_z)$, from our existing diffusion coefficients in velocity. The derivation is given in Section~S2 of the Supplementary Material. For fixed $\theta_z$, we have,
\begin{align}
     D^{(1)}_J(J_z | \theta_z) =  \frac{v_z(J_z,\theta_z)}{\Omega_z} D^{(1)}_v(v_z) + \frac{1}{2\Omega_z }D^{(2)}_{vv}(v_z),
\end{align}
and,
\begin{align}
    D^{(2)}_{JJ}(J_z | \theta_z) = \frac{v_z^2}{\Omega_z^2} D^{(2)}_{vv}(v_z),
\end{align}
where $z=z(J_z,\theta_z)$ and $v_z=v_z(J_z,\theta_z)$. The orbit averages are then obtained by integrating these expressions over $\theta_z$, with the spatial dependence of the GMC layer included through the local density dependence of the velocity diffusion coefficients. This is explained and derived in Section~S2 of the Supplementary Material.
\par
Finally, we set $\partial f/\partial t = 0$ in equation (\ref{eq:FP-action-orbit-average}), where the flux becomes constant and we take it to be $0$ as in equation (\ref{eq:ss-velocity}), which yields,
\begin{align} \label{eq:fJ-SS}
    f_z(J_z) \propto \frac{1}{\overline{D^{(2)}_{JJ}}(J_z)} \, \exp \left( \int^{J_z} \frac{2 \overline{D^{(1)}_J}(J_z')}{\overline{D^{(2)}_{JJ}}(J_z')} \,  dJ_z' \right).
\end{align}
This is the equilibrium DF associated with a fixed scattering environment. It should not be interpreted as the asymptotic state of the fully time-dependent problem, because in the latter case the in-plane velocity distribution and the GMC density evolve with time. To plot equilibrium action distributions in equation (\ref{eq:fJ-SS}), we must numerically integrate the orbit-averaged diffusion coefficients in action, as described in Section~S2 of the Supplementary Material. Finally, the integrand of equation (\ref{eq:fJ-SS}) is integrated numerically with a cumulative trapezoidal rule. 

\begin{figure}
    \centering
\includegraphics[width=\linewidth]{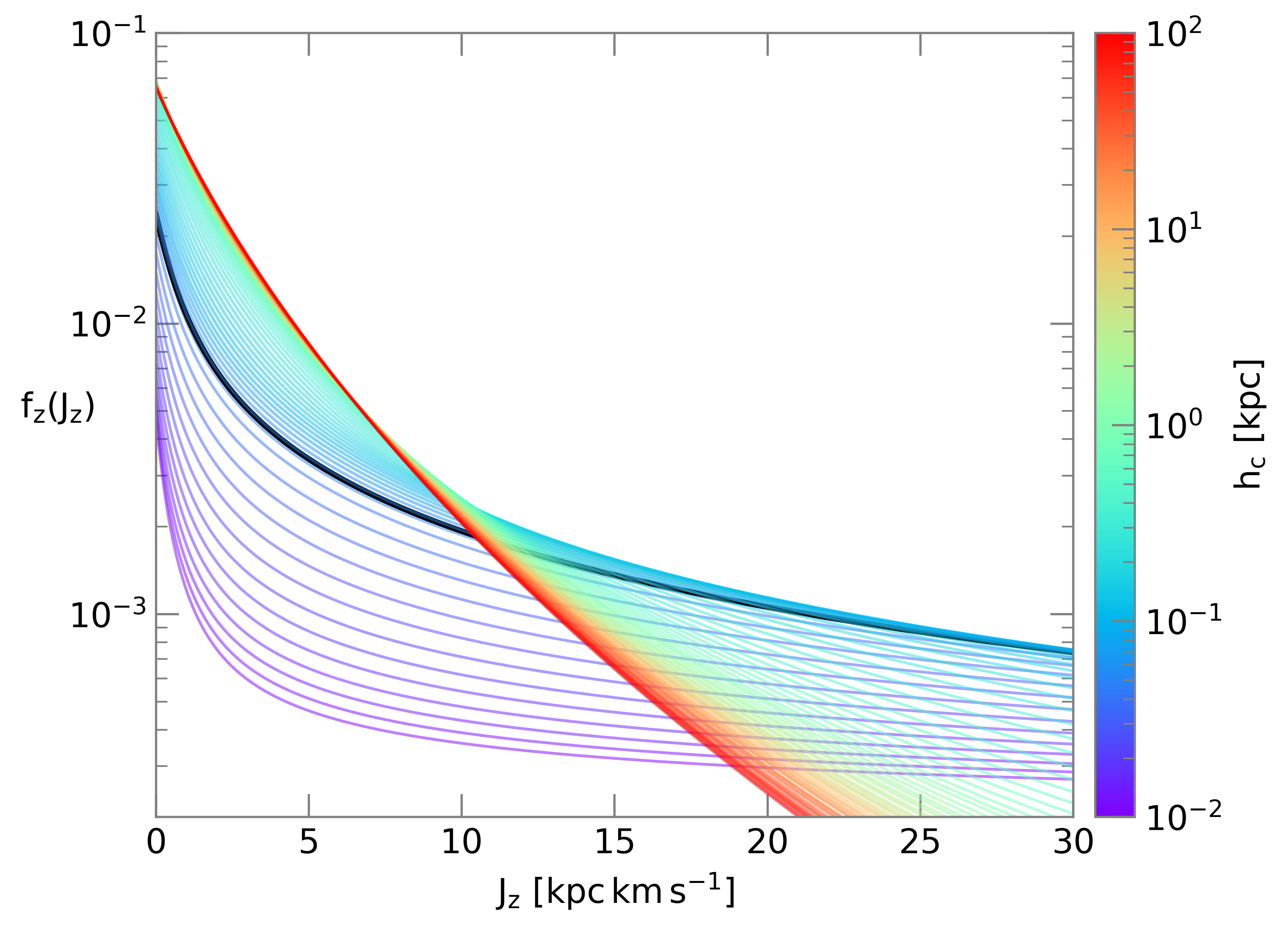}
    \caption{Steady state action distributions computed numerically using equation (\ref{eq:fJ-SS}). The curves are coloured by different scaleheights of the GMC cloud distribution, $h_c$. The black curve denotes the distribution for scaleheight, $h_c=0.05$ kpc. }
    \label{fig:f-jz-all-hc}
\end{figure}
\par
\begin{figure*}
\centering
\includegraphics[width=\textwidth]{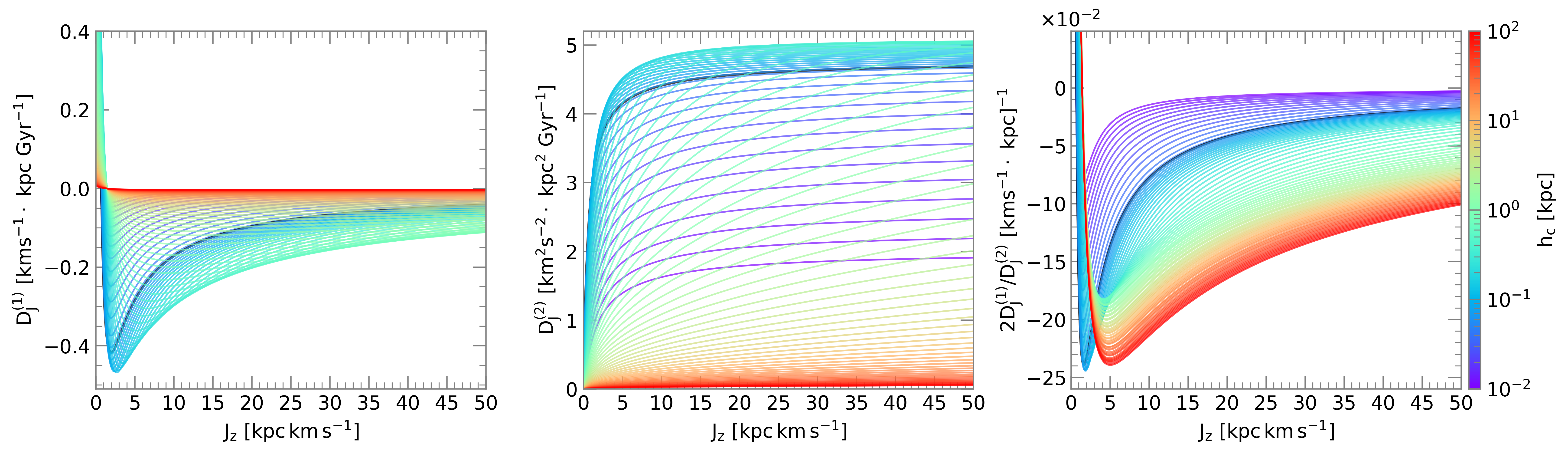}
    \caption{\textit{Left panel}: First orbit-averaged action diffusion coefficient, $\overline{D^{(1)}_J}(J_z)$. \textit{Middle panel}: Second orbit-averaged action diffusion coefficient, $\overline{D^{(2)}_{JJ}}(J_z)$. \textit{Right panel}: Ratio of the diffusion coefficients which appear in the integrand of the exponent of the stationary solution in equation (\ref{eq:fJ-SS}). In all panels, curves are coloured by scaleheight of the GMC distribution, $h_c$. The black curve denotes the scaleheight, $h_c=0.05$ kpc. }
    \label{fig:dj-coeffs}
\end{figure*}

Figure \ref{fig:f-jz-all-hc} displays the stationary solutions $f_z(J_z)$ as per equation (\ref{eq:fJ-SS}), coloured by different scaleheights of the GMC cloud $h_c$. Evidently, as $h_c \rightarrow 0$, the distribution tends to a cusp at small actions, and as $h_c \rightarrow \infty$, the distribution tends to the pseudo-isothermal case. The latter corresponds to scattering randomised in all of phase-space, rather than being concentrated in the midplane. Consequently, the narrow midplane distribution of GMCs is one contributing factor to a cuspy distribution at small actions. Moreover, Figure \ref{fig:dj-coeffs} displays (from left to right), the first and second orbit averaged diffusion coefficients, as well as the ratio of the two, which appears in the integrand of the exponent in the steady state solution. The middle panel in particular is an indicator of why we get a cusp at small actions. Namely, the reciprocal of the second order diffusion coefficient appears in the stationary solution (see equation (\ref{eq:fJ-SS})) and thus a strongly non linear relation at small actions is what steepens the action distribution here. A non-constant ratio of the diffusion coefficients also contributes to a non-linear exponent in the stationary solution. We will discuss the implications of the shape of the diffusion coefficients in a forthcoming paper. In the limit of large actions, the ratio tends to asymptote, which accounts for near-isothermality in this regime, as seen by approximately exponential behaviour at large actions in Figure \ref{fig:f-jz-all-hc}. We remark that the orbit-averaged coefficients are evaluated in the harmonic limit, assuming a constant frequency, $\Omega_z$, and the mappings $z(J_z,\theta_z) = \sqrt{2J_z/\Omega_z}\cos \theta_z$ and $v_z(J_z,\theta_z) = \sqrt{2J_z\Omega_z}\sin \theta_z$. However, as discussed in Section S2 of the Supplementary Material, this approximation is well motivated for realistic GMC layers, which are concentrated near the midplane, precisely where the potential is near harmonic. Therefore, for large $h_c$, we expect the equilibrium DFs presented here to be valid only in a harmonic potential. Finally, we do not plot results for $h_c \gtrsim 100$ kpc, as the results do not vary for large scaleheights.

\section{A new stationary DF} \label{sect:new-df}
\par
In light of the analysis of our equilibrium distributions, for frozen-in diffusion coefficients, we seek a generalised vertical action distribution, to bracket the two limiting behaviours being: the razor-thin and infinitely extended GMC distribution. Taking the logarithmic derivative of equation \eqref{eq:fJ-SS} yields,
\begin{align} \label{eq:slope-func-analytic}
    \frac{d\ln f}{dJ_z}=\frac{2D_J^{(1)}}{D_{JJ}^{(2)}}-\frac{d\ln D_{JJ}^{(2)}}{dJ_z}.
\end{align}
This quantity is more appropriate for fitting than $f(J_z)$ itself, as it directly probes the shape of drift and diffusion coefficients. Despite non-monotonicity in the ratio of diffusion coefficients, the combination in the slope function of equation \eqref{eq:slope-func-analytic} is monotonically increasing, and was empirically chosen to take the form: 
\begin{align}
    \frac{d\ln f}{dJ_z}
    =
    -\frac{\alpha}{2J_c}
    \left(1+\frac{J_z}{J_c}\right)^{-\gamma}
    -
    \frac{\Omega_0/\sigma_z^2}
    {1+\Omega_0 J_z/(\eta\sigma_z^2)},
    \label{eq:slope_model}
\end{align}
where $\Omega_0$ is the small-amplitude (harmonic) vertical frequency, $\sigma_z$ sets the characteristic low-action velocity scale, $\eta$ controls the high action power-law tail, and $\alpha$, $J_c$ and $\gamma$ describe the correction arising from the extended GMC cloud. Then $J_s\equiv \sigma_z^2/\Omega_0$ sets the approximately exponential action scale.

We interpret the scale \(J_c\), as the action at which an orbit begins to leave the GMC layer. In the harmonic limit, \(J_z=\Omega_0 z_{\max}^2/2\), so an orbit with vertical amplitude comparable to the GMC scaleheight, \(z_{\max}\sim h_c\), has \(J_z\sim \Omega_0 h_c^2\), up to a numerical factor. We therefore expect \(J_c \sim \Omega_0 h_c^2\). In the fits below, however, \(J_c\) is left free for each \(h_c\), and we then compare the fitted trend with this geometric expectation. Physically, \(J_c\) marks the action below which stars remain mostly embedded in the GMC layer, and above which stars spend an increasing fraction of their orbit outside the scattering layer. Equation~\eqref{eq:slope_model} integrates analytically to,
\begin{align}
    f(J_z)\propto\exp\left[-\frac{\alpha}{2}
        H_\gamma\left(\frac{J_z}{J_c}\right)\right]
    \left(1+\frac{\Omega_0 J_z}{\eta\sigma_z^2}
    \right)^{-\eta},
    \label{eq:df_model}
\end{align}
where,
\begin{align}
    H_\gamma(x)=\int_0^x (1+u)^{-\gamma}\,du=
    \begin{cases}
        \ln(1+x), & \gamma = 1, \\[6pt]
        \dfrac{(1+x)^{1-\gamma}-1}{1-\gamma},
        & \gamma \neq 1 .
    \end{cases}
    \label{eq:Hgamma}
\end{align}
\par
While the slope function was chosen rather heuristically, the DF has a useful physical interpretation. The second factor, in the limit \(\eta\rightarrow\infty\), reduces to the isothermal form,
\begin{align}
    \left(
        1+\frac{\Omega_0 J_z}{\eta\sigma_z^2}
    \right)^{-\eta}
    \longrightarrow
    \exp\left(
        -\frac{\Omega_0 J_z}{\sigma_z^2}
    \right),
\end{align}
and this term was inspired by the rational linear distribution function (RLDF) in \citet{Li_Widrow_2021}, which was also later written in terms of vertical action in \citet{Gilman_2025}. For \(\eta\rightarrow\infty\), $\sigma_z$ represents the velocity dispersion in the isothermal limit. We introduce the additional prefactor in equation \eqref{eq:df_model}, which we call the \textit{GMC-layer correction}, to account for the fact that scattering is weighted by the vertical distribution of GMCs rather than being uniform along the orbit.
\par
Let us now consider the two limiting cases. Firstly, in the razor-thin GMC limit, we have \(h_c\rightarrow 0\), which implies, \(J_c\rightarrow 0\). So, for any fixed non-zero action \(J_z\) we have \(J_z/J_c\gg 1\). To recover the cusp associated with the razor-thin GMC layer, we take the limiting value, $\gamma=1$, so that
\begin{align}
    \exp\left[
        -\frac{\alpha}{2}
        H_1\left(\frac{J_z}{J_c}\right)
    \right]
    =
    \left(
        1+\frac{J_z}{J_c}
    \right)^{-\alpha/2}
    \simeq
    \left(
        \frac{J_z}{J_c}
    \right)^{-\alpha/2}.
\end{align}
Further, taking the limit $\eta \rightarrow\infty$, recovers the exponential tail, so that equation \eqref{eq:df_model} becomes,
\begin{align}
    f_z(J_z) \propto
    J_z^{-\alpha/2}
    \exp\left(
        -\frac{\Omega_0 J_z}{\sigma_z^2}
    \right),
\end{align}
which recovers the form in the razor-thin limit in \S \ref{sect: razor-thin-gmc}, by choosing $\alpha=1$ for the harmonic oscillator, and $\Omega_0\equiv \Omega_z$ is the vertical frequency. In the linear potential, since the orbital frequency is non-constant, we must interpret the combination, $\frac{\sigma^2_z}{\Omega_0} \equiv J_0$ as the action scale, and not $\Omega_0$ as the true orbital frequency. A choice of $\alpha=2/3$ then recovers the razor-thin GMC limit, in this linear potential. It should be remarked that in \S \ref{sect: razor-thin-gmc}, we assumed a \textit{constant} diffusion, so these limiting values for $\eta$ and $\gamma$ need not impose bounds for the equilibrium DFs with an extended GMC distribution, as we may expect the parameters to vary differently, with velocity dependent scattering. We merely show that our empirical DF can recover this simple case for constant diffusion coefficients. 
\par
Now, consider the opposite limit, \(h_c\rightarrow\infty\), which corresponds to GMC scattering randomised in all of phase space, rather than being concentrated in the midplane. Since we expect \(J_c \sim \Omega_0 h_c^2\), this corresponds to \(J_c\rightarrow\infty\).
For any finite \(J_z\), $\frac{J_z}{J_c}\rightarrow 0$, and therefore $H_\gamma$, to first order can be written as, 
\begin{align}
    H_\gamma\left(\frac{J_z}{J_c}\right)
    =
    \frac{J_z}{J_c}
    +
    \mathcal{O}\left[
        \left(\frac{J_z}{J_c}\right)^2
    \right],
\end{align}
so that,
\begin{align}
    \exp\left[
        -\frac{\alpha}{2}
        H_\gamma\left(\frac{J_z}{J_c}\right)
    \right]
    \longrightarrow 1.
\end{align}
In this limit the GMC-layer correction disappears, and the distribution reduces to,
\begin{align} \label{eq:thick_limit_powerlaw}
    f_z(J_z)
    =
    C_\infty
    \left(
        1+\frac{\Omega_0 J_z}{\eta\sigma_z^2}
    \right)^{-\eta}.
\end{align}
The normalisation condition gives,
\begin{align}
    C_\infty
    =
    \frac{\Omega_0}{2\pi\sigma_z^2}
    \frac{\eta-1}{\eta},
    \qquad
    \eta>1.
\end{align}
In the further limit, \(\eta\rightarrow\infty\), equation~\eqref{eq:thick_limit_powerlaw} becomes
\begin{align}
    f_z(J_z)
    =
    \frac{\Omega_z}{2\pi\sigma_z^2}
    \exp\left(
        -\frac{\Omega_z J_z}{\sigma_z^2}
    \right),
\end{align}
which is the familiar pseudo-isothermal DF, where we assume $\Omega_0 = \Omega_z$ is the harmonic frequency. 
\par
Further, for a finite velocity dispersion, we require $\eta >2$, if $\gamma>1$. In the special case where \(\gamma=1\), then,
\begin{align}
    \exp\left[
        -\frac{\alpha}{2}
        H_1\left(\frac{J_z}{J_c}\right)
    \right]
    =
    \left(
        1+\frac{J_z}{J_c}
    \right)^{-\alpha/2},
\end{align}
and hence at large actions, $f_z(J_z)\sim J_z^{-(\eta+\alpha/2)}$. Then, we require $\eta+\frac{\alpha}{2}>1$ for normalisability, and $\eta+\frac{\alpha}{2}>2$ for a finite velocity dispersion. 

\subsection{Fitting procedure}
\par
\begin{figure}
    \centering
    \includegraphics[width=\linewidth]{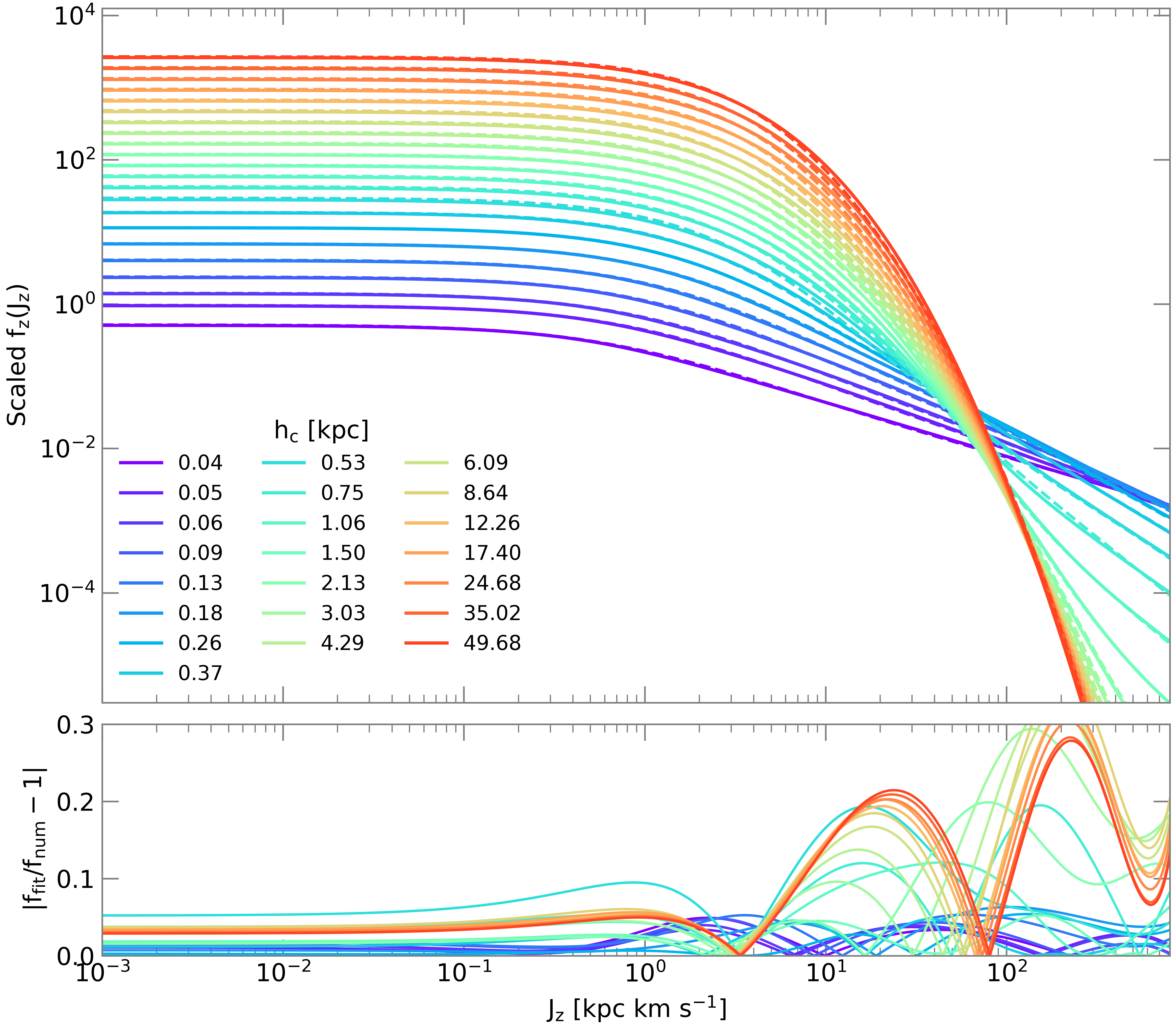}
    \caption{Fits of the empirical DF family in equation~\eqref{eq:df_model} to the numerical zero-flux equilibrium solutions. The upper panel shows the numerical DFs, scaled in amplitude for visualisation, with corresponding best-fitting curves for a range of GMC scaleheights, $h_c$, shown with dashed lines. The lower panel shows the fractional residuals, $|f_{\rm fit}/f_{\rm num}-1|$, where $f_{\rm fit}$ is the fitted curve, and $f_{\rm num}$ is the numerical solution.}
    \label{fig:family-fit-curves}
\end{figure}

\begin{figure}
    \centering
    \includegraphics[width=\linewidth]{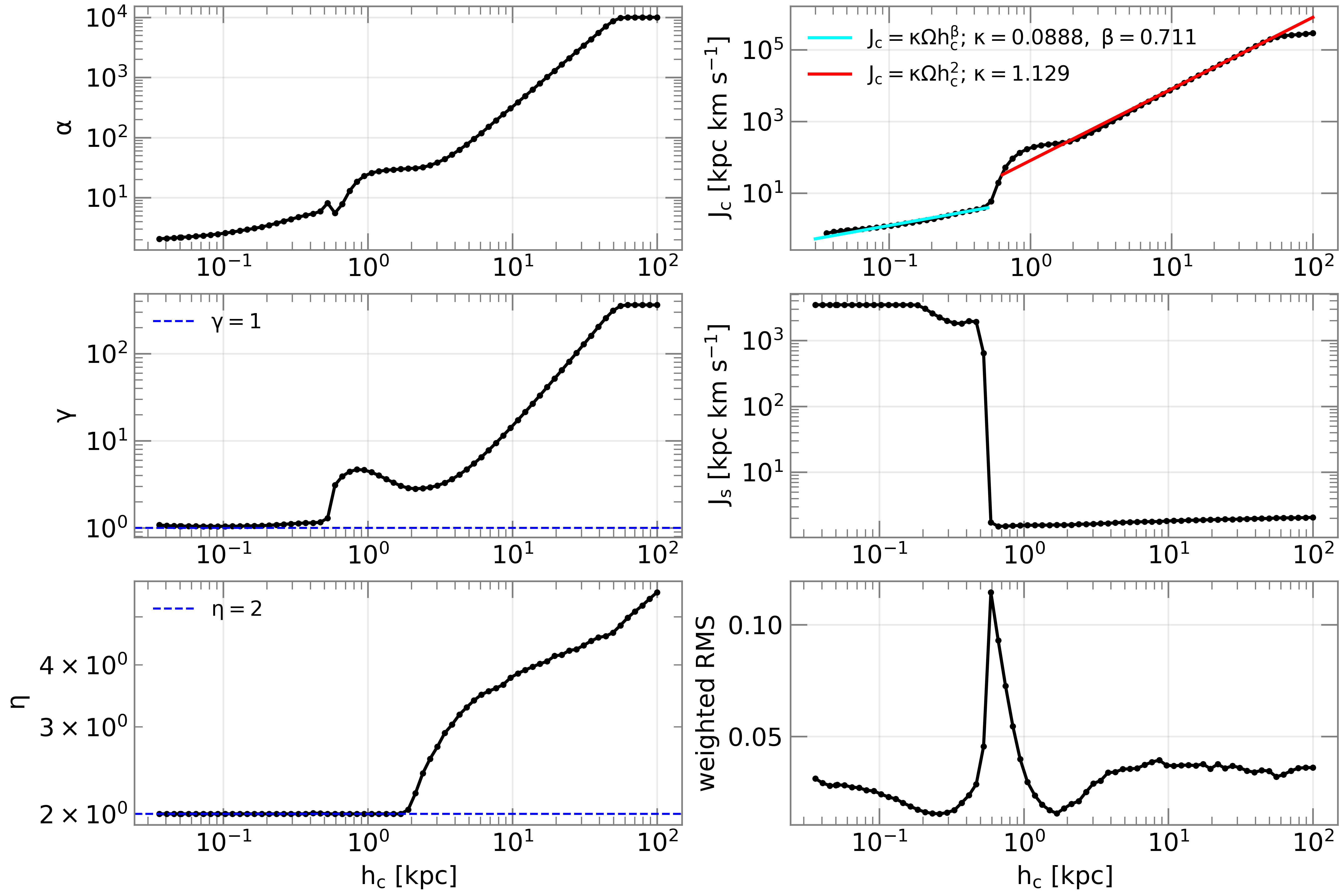}
    \caption{Best-fitting parameters of the empirical DF family as functions of GMC scaleheight, $h_c$. The panels show $\alpha$, $J_c$, $\gamma$, $J_s \equiv \sigma^2/\Omega_0$, $\eta$, and the weighted root-mean-square (RMS) fit error. The transition action $J_c$ follows two approximate regimes. At small $h_c$ it varies as \(J_c\propto h_c^{0.711}\), shown by the best-fitting cyan line, while at large \(h_c\) it approaches \(J_c\simeq1.129\,\Omega_0h_c^2\), indicated by the red line. The dashed horizontal lines in the $\gamma$ and $\eta$ panels indicate the reference values $\gamma=1$ and $\eta=2$, respectively.}
    \label{fig:family-fit-params}
\end{figure}

To investigate how the shape of the equilibrium DFs varies with cloud scaleheight, $h_c$, we fit the empirical DF family to the numerical zero-flux solutions by fitting the logarithmic slope rather than the DF itself. We fit,
\begin{align}
    p_{\rm eff}(J_z; h_c)\equiv
-J_z\frac{d\ln f_z}{dJ_z},
\end{align}
where $d\ln f_z/dJ_z$ is given in equation \eqref{eq:slope_model}. Fitting $p_{\rm eff}(J_z)$ weights the inner and outer action ranges more evenly than fitting $d\ln f_z/dJ_z$ directly. The full family of \(h_c\) curves is fitted simultaneously on the interval $0 \leq J_z \leq 800$ kpc km s$^{-1}$, using a weighted least-squares fit that gives additional weight to the cusp and tail regions.
\par
The top panel of Figure~\ref{fig:family-fit-curves} displays the empirical DF fit (dashed) to the numerical zero-flux equilibrium solutions (solid) for selected scaleheights, where the amplitudes of the DFs are scaled to separate the fits visually for the various scaleheights. The bottom panel reveals a strong agreement between the fitted and numerical curves, with fractional residuals within $\sim 0.05$ for the smallest scaleheights over the fitted range, while errors are larger for the largest scaleheights. For the thickest GMC layers, the curves approach the pseudo-isothermal form, and so the GMC-layer correction term becomes weakly constrained. However, our model is intended to capture the change in DF shape induced by the finite vertical extent of the GMC layer, and fits the numerical curves sufficiently well at realistic GMC scaleheights of up to $\sim 100-200$ pc. 
\par
The fitted parameters, shown in Figure~\ref{fig:family-fit-params}, as functions of $h_c$ reveal two regimes. For small GMC scaleheights, roughly $h_c\lesssim 0.1$--$0.2\,{\rm kpc}$, the parameters $\alpha$, $\gamma$, and $\eta$ vary only weakly. In this regime, the primary dependence on $h_c$ enters through the GMC-layer scale $J_c$, while the shape of the correction is approximately fixed. In particular, $\gamma\simeq1$, so the finite-thickness factor behaves approximately as a power law,
\begin{align}
    \exp\left[
        -\frac{\alpha}{2}
        H_\gamma\left(\frac{J_z}{J_c}\right)
    \right]
    \simeq
    \left(1+\frac{J_z}{J_c}\right)^{-\alpha/2},
\end{align}
which has the same analytic form as the razor-thin analytic solutions in \S\ref{sect: razor-thin-gmc}, although the fitted exponent is not expected to equal the values obtained assuming constant diffusion coefficients. For example, at $h_c=0.05\,{\rm kpc}$ we find, $\alpha=2.168$, $\gamma = 1.049$, $\eta = 2.000$ and $J_s\simeq3.5\times10^3\,{\rm kpc\,km\,s^{-1}}$, which corresponds to $\sigma_z \sim 500\,{\rm km\,s^{-1}}$ for a choice of $\Omega_0 = 72$ km s$^{-1}$ kpc$^{-1}$. The effective low-action power-law correction has exponent $\alpha/2\simeq 1.0$, which is steeper than the constant-kick harmonic razor-thin GMC value of $1/2$. This should be interpreted as a consequence of the more realistic, velocity-dependent scattering, not as an inconsistency with the analytic limiting case. 
\par
Moreover, using $J_s\equiv \frac{\sigma_z^2}{\Omega_0}$, the RLDF-like factor can be written as,
\begin{align}
    \left(1+\frac{J_z}{\eta J_s}\right)^{-\eta}.
\end{align}
Thus $J_s$ sets the approximately exponential action scale, while $\eta J_s$ sets the turnover to the asymptotic power-law tail. In the thin-layer fits, $J_s$ lies well beyond the fitted action range, so the fit does not directly measure the high-action tail scale. The large inferred values of $\sigma_z^2 = \Omega_0 J_s$ should therefore not be read as physical vertical velocity dispersions. Rather, they indicate that the DF shape over the fitted range is dominated by the GMC-layer correction. At the smallest scaleheights, the fitted $J_c$ does not follow the simple quadratic scaling $J_c\propto \Omega_0 h_c^2$. Instead, over the resolved thin-layer range it is better described by a shallower scaling as shown by the cyan line in Figure \ref{fig:family-fit-params}. We do not interpret this as evidence for a different physical transition action. In the limit of a very thin GMC layer, $J_c$ lies below much of the fitted action interval, and
\begin{align}
    \left(1+\frac{J_z}{J_c}\right)^{-\alpha/2}
    \simeq
    J_c^{\alpha/2}J_z^{-\alpha/2}.
\end{align}
The dependence on $J_c$ is then partly absorbed into the normalisation, leaving the logarithmic slope more sensitive to the effective cusp exponent than to the precise value of the transition action. Hence $J_c$ should be regarded as an empirical transition scale in the very thin-layer limit, while $J_c\sim \Omega_0h_c^2$ remains the expected geometric scaling when the transition is well resolved.

In the thin GMC-layer regime, we find useful fiducial values,
\begin{align}
    \gamma \simeq 1,\qquad
    \eta \simeq 2,\qquad
    \alpha \simeq 2.0\text{--}2.5,
\end{align}
with the remaining dependence on cloud thickness entering mainly through $J_c(h_c)$ and the overall normalisation. This reduced parametrisation should not be extrapolated to arbitrarily large $h_c$, where the GMC-layer correction becomes degenerate and the harmonic approximation used in the orbit averaging is less well motivated. For thin layers, we expect the DF with these parameter values to be insensitive to the choice of potential, given the cloud scaleheights are small. 
\par
Conversely, for large $h_c$, the fitted behaviour changes qualitatively. The transition action scales as $J_c\simeq \Omega_0 h_c^2$, as shown in the top right panel of Figure \ref{fig:family-fit-params}. Over any finite fitted action interval, this implies $J_z/J_c\ll1$, so
\begin{align}
    H_\gamma\left(\frac{J_z}{J_c}\right)
    \simeq
    \frac{J_z}{J_c},
\end{align}
and the finite-thickness prefactor tends to unity, up to a weak residual slope proportional to $\alpha/J_c$. In this regime the individual values of $\alpha$ and $\gamma$ become highly degenerate and should not be assigned direct physical meaning. Their rapid growth in Figure~\ref{fig:family-fit-params} reflects the fact that the model is trying to fit a correction that is already disappearing from the observable action range.
\par
The parameter $\eta$ increases with $h_c$, showing that thicker GMC distributions drive the stationary DF toward an exponential tail. Since $\eta\rightarrow\infty$ gives the pseudo-isothermal limit, this trend is consistent with the interpretation that a vertically extended scattering population randomises the diffusion more uniformly along the orbit. At the same time, $J_s$, and therefore $\sigma_z$, drops from the large, weakly constrained values found in the thin GMC-layer regime and approaches the velocity scale of the nearly pseudo-isothermal equilibrium. Thus $\sigma_z$ becomes physically meaningful only as the GMC-layer correction becomes small. This is contrary to the thin-layer regime, where it is mainly a nuisance parameter controlling a tail that lies outside the fitted range.

Finally, the large-$h_c$ fits should be treated as a useful limiting experiment rather than as a realistic description of GMC scattering in the Galactic disc. The orbit-averaged coefficients were computed using the harmonic mappings associated with the small-amplitude frequency $\Omega_0$. This is well justified for realistic GMC layers, because most scattering occurs close to the midplane where the vertical potential is approximately harmonic. If the cloud layer is artificially extended to very large $h_c$, however, the scattering samples heights where a realistic Galactic potential is anharmonic. In that regime a fully self-consistent calculation would require the true functions $z(J_z,\theta_z)$, $v_z(J_z,\theta_z)$, and $\Omega_z(J_z)$, rather than the harmonic approximation. Our main physical conclusions are therefore drawn from the small-$h_c$ regime, where the GMC layer is thin and the harmonic approximations are most reliable.

\section{Time-dependent evolution} \label{sect:new-test-particle-sim}
\par
While the stationary solutions are theoretically novel, the relaxation time to such a state can exceed $\sim 100$ Gyr, particularly at large actions. Here, we perform a more realistic test-particle simulation using MW parameters, and allow time-dependence of velocity moments, to show that the functional form of the DF in equation \eqref{eq:df_model} can be extended for more realistic time-dependent systems. This begins by reintroducing the decaying GMC density as per equation (\ref{eq:gmc-with-decay}), and modelling in-plane heating which enters via the velocity moments (\textit{see} Appendix \ref{app: relative speed} and \ref{app:moments-derivation} for more details). Furthermore, we validate this with direct numerical solutions of the Fokker-Planck equation with time-dependent coefficients.

\subsection{GMC distribution} 
\par
We consider a cloud layer with a Gaussian number density centred at the Galactic midplane, given by,
\begin{align} \label{eq:cloud-density}
    n_c(z) = 
        n_0 \, \exp [-z^2/(2h_c^2)]
\end{align}
where $h_c = 0.05$ kpc is chosen as a realistic MW value in the Solar neighbourhood, \citep{Nakanishi_2006}.
\par
It is known that GMC heating is more efficient earlier on \citep{ABS_2016}, when star formation rates (SFR) are higher, so we incorporate a declining SFR and GMC density over time. We model the SFR as,
\begin{equation}
    \text{SFR}(t) \propto \exp(-t/t_{\mathrm{SFR}}),
\end{equation}
where $t_{\mathrm{SFR}} = 8$ Gyr. To estimate the evolution of GMC density with time, we adopt the approach of \citet{Ting_Rix_2019}, who assume that $\Sigma_{\text{GMC}}$ scales with $\Sigma_{\text{SFR}}$ via the Schmidt–Kennicutt law as $\Sigma_{\text{GMC}} \sim \Sigma_{\text{SFR}}^{1/\alpha_{\text{KS}}}$ \citep{Kennicutt_Evans_2012}. We choose $\alpha_{\text{KS}} = 1.0$, the index measured for cold molecular clouds \citep{Bigiel_2008}. Hence, the GMC surface density is directly proportional to the SFR density. The spatial GMC density is convolved with a function of time, so that,
\begin{align} \label{eq:gmc-with-decay}
    n_c(z,t) = 
        n_0 \, \exp [-z^2/(2h_c^2)] \, \exp(-t/t_{\mathrm{GMC}}),
\end{align}
where $t_{\mathrm{GMC}} = 8$ Gyr. We determine $n_0$ so that $n_c(z \, | \, t=10 \, \rm{ Gyr})$ matches a present-day midplane molecular GMC mass density of $\Sigma_{\rm{GMC}} \approx 5$ M$_\odot$ pc$^{-2}$ \citep{Sharma_2021}. 
Hence, the midplane mass density at present day is estimated as $\rho_0 \approx 0.039$ M$_\odot$ pc$^{-3}$. For a total run time of $T=10$ Gyr, the initial midplane density must be $\sim \exp(1.25)$ times higher. Finally, $n_0$ is the initial midplane mass density normalised by a typical GMC cloud mass, $\langle M_c \rangle$. Based on measurements of molecular gas in the Galaxy \citep{Nakanishi_2006,Kokaia_2019}, both the vertical and radial density of GMCs in the MW are variable but we exclude a radial dependence here due to the one-dimensional nature of the model, and aim to study the behaviour in the Solar neighbourhood only. Hence, we assume a constant scaleheight for the GMC cloud layer in equation (\ref{eq:gmc-with-decay}).
\par
GMC masses are drawn from a power-law mass function, $dN/dM_c \propto M_c^\gamma$, with $\gamma = -1.6$ \citep{ABS_2016, Jeffreson_2022}. We set mass limits to $M_{\text{lower}} = 10^5 $ M$_\odot$ and $M_{\text{upper}} = 2.6 \times 10^6 $ M$_\odot$, as larger GMCs tend to dominate scattering effects, and there are fewer clouds in the MW with masses over $\sim 10^7$ M$_\odot$. This gives a mean mass of $\langle M_c \rangle \approx 10^6$ M$_\odot$.

\subsection{Test particle simulations} \label{sect:test-particle-sim-time}
\par
We simulate directly the SDE in equations (\ref{eq:sde1}) and (\ref{eq:sde2}), which are discrete stochastic evolutions of the Fokker-Planck equation (\ref{eq:FP-velocity}), using an Euler-Maruyama (EM) discretisation. Due to state-dependent drift and diffusion coefficients that are strongly non-linear, a fixed-timestep EM scheme can be unstable or biased, as the coefficients are non-globally Lipschitz. \citet{Fang_Giles_2016} prove strong convergence and stability of an adaptive EM scheme, when the timestep is a state-dependent function, $h(x)$ that is (i) bounded from above, (ii) shrinks as the state enters stiffer regions and (iii) is monotonic. In light of this, we choose an adaptive timestep to be a function of action,
\begin{align}
    \Delta t(J_z) = \eta \, t_{\rm cross} (J_z),
\end{align}
where $\eta=0.05$, but is tuneable, and $t_{\rm cross} (J_z)$ is the cloud-crossing time. Thus, before imposing the timestep floor, $\eta=0.05$ corresponds to roughly 20 substeps per cloud crossing. For a harmonic oscillator potential, with frequency, $\Omega_z$, this can be expressed analytically as 
\begin{align} \label{eq:t_cross}
    t_{\rm cross} (J_z) = \begin{cases}
        \frac{2}{\Omega_z} \sin^{-1}(h_c /z_{\rm max}), \quad &z_{\rm max} \geq h_c, \\
        \frac{\pi}{\Omega_z}, & z_{\rm max} < h_c,
    \end{cases}
\end{align}
where $h_c$ is the cloud scaleheight, and $z_{\rm max} = \sqrt{2J_z/\Omega_z}$, and equation (\ref{eq:t_cross}) satisfies conditions (i)-(iii) above. In the thin cloud layer approximation, using the midplane speed, $v_0 = \sqrt{2\Omega_z J_z}$
\begin{align}
    t_{\rm cross} (J_z) \approx \frac{2h_c}{v_0} = \frac{2h_c}{\sqrt{2\Omega_z J_z}} \propto J_z^{-1/2},
\end{align}
and thus the timestep shrinks as $\sim J_z^{-1/2}$ as the star passes the cloud layer at a larger velocity, exactly where we need a higher resolution to avoid skipping encounters. This choice of adaptive timestep ensures strong 1/2-order convergence and stability, for a finite time-interval \citep{Fang_Giles_2016}. The timestep is already bounded from above, and we further bound from below with $\Delta t_{\rm min}$ being one RK4 timestep in the orbit integrator, which is set to $\Delta t = 0.179$ Myr. For non-harmonic potentials we adopt the same time-stepping regime, using a local frequency, given that for small $|z| \lesssim h_c$, any smooth disc potential is approximately harmonic with $\Omega_z \approx \partial ^2\Phi /\partial z^2 |_{z=0}$.

\par
We choose code normalisations, $G=1$, $R_0 = 8.2$ kpc, $V_0= 229.05$ km s$^{-1}$, $M_0 = 10^{11}$M$_{\odot}$ so that one time unit is $T_0 = 35.8$ Myr. We evolve in the harmonic oscillator potential, with $\Omega_z = 72$km s$^{-1}$kpc$^{-1}$, as well as the isothermal slab potential, $\Phi(z) = 2\sigma_{\rm slab}^2 \log \cosh[\frac12 z/z_0]$, with $\sigma_{\rm slab}=21.65$ km s$^{-1}$ and $z_0 = 0.23$ kpc.
\par
We sample $N=10^5$ stars initially from an exponential density $\rho(z) \propto \text{exp}[-|z|/h]$, where we set $h = 60$ pc; this is a pragmatic choice to avoid starting at $z=0$. Since we will study the effects of GMC scattering on Gyr timescales, the distributions at later times are insensitive to this initial profile. Given the birth (vertical) velocity dispersion of stars is $\sim 6$ km s$^{-1}$ \citep{ABS_2016,Sharma_2021}, we sample initial velocities from a normal distribution, $\mathcal{N}(0,\sigma_{z,0})$, where we set $\sigma_{z,0} = 6$ km s$^{-1}$. This is representative of a thin-disc population. For this simulation, we birth all stars at $t=0$, so that age $\tau=t$ equals time, to investigate the time-evolution of a single coeval population. Unless otherwise stated, we keep the GMC scaleheight $h_c = 0.05$ kpc, cloud velocity dispersion, $\sigma=6$ km s$^{-1}$ in equation (\ref{eq:cloud-maxwellian}), the present-day molecular GMC surface density, $\Sigma_{\rm GMC} = 5$ M$_{\odot}$ pc$^{-2}$ and the GMC density timescale, $t_{\mathrm{GMC}} = 8$ Gyr, in equation (\ref{eq:gmc-with-decay}), fixed. 

\subsection{Numerical solution of the time-dependent Fokker--Planck equation}
\label{subsec:numerical-fp-J}
\par
For the harmonic oscillator potential, we solve the orbit-averaged Fokker--Planck equation numerically in vertical action. We begin with the time-dependent equation in flux form, given by equations \eqref{eq:fp-flux-form}-\eqref{eq:flux-term}. The time-dependent coefficients are constructed from precomputed velocity-space moment tables, as described in Appendix \ref{app:moments-derivation}, where in-plane heating and the time-dependent GMC distribution are included, with a cloud density decay timescale, $\tau_{\rm GMC}=8$ Gyr, as in the test-particle simulations. The orbit-averaged coefficients are then computed as described in Section S2 of the Supplementary Material. The resulting arrays $D_J^{(1)}(J_z,t)$ and $D_{JJ}^{(2)}(J_z,t)$ are tabulated on the same fixed $J_z-$grid used for the Fokker--Planck evolution.
\par
We discretise equation \eqref{eq:FP-action-orbit-average} on a fixed, logarithmically spaced grid in $J_z$. The numerical solution $f_i(t)$ represents the cell-averaged distribution in the interval $J_{i-1/2}<J_z<J_{i+1/2}$. Integrating over each cell gives the finite-volume evolution equation,
\begin{align}
\frac{d f_i}{dt}
&=-\frac{1}{\Delta J_i}\left[F_{i+1/2}-F_{i-1/2}\right],
\label{eq:finite-volume-fp}
\end{align}
where $\Delta J_i=J_{i+1/2}-J_{i-1/2}$. The flux, $F_{i\pm 1/2}$, is split into a drift term and a diffusive term. The drift term, \(D_J^{(1)}f\), behaves like advection in action space. If \(D_J^{(1)}>0\), probability flows toward larger \(J_z\), so the value of \(f\) at the interface \(J_{i+1/2}\) is taken from the left cell. If \(D_J^{(1)}<0\), probability flows toward smaller \(J_z\), so the value of \(f\) is taken from the right cell. At each cell interface, we take \(f\) from the cell on the side from which the probability is flowing. Thus, at the interface \(J_{i+1/2}\),
\begin{align}
f_{\rm up}
&=
\begin{cases}
f_i, & D_{J,i+1/2}^{(1)} \ge 0,\\
f_{i+1}, & D_{J,i+1/2}^{(1)} < 0.
\end{cases}
\end{align}
The diffusive part of the flux is evaluated with a centred finite
difference of the quantity \(D_{JJ}^{(2)}f\). The total numerical flux is then
\begin{align}
F_{i+1/2}
&=
D_{J,i+1/2}^{(1)} f_{\rm up}
-
\frac{1}{2}
\frac{
D_{JJ,i+1}^{(2)} f_{i+1}
-
D_{JJ,i}^{(2)} f_i
}{
J_{i+1}-J_i
}.
\label{eq:numerical-flux-fp}
\end{align}

This form preserves the conservative structure of the Fokker--Planck equation, because the flux through each cell boundary is removed from one cell and added to the neighbouring cell. We impose zero-flux boundary conditions, at lower and upper edges of the domain, $F_{1/2}=0$ and $F_{N+1/2}=0$, which prevent probability entering or leaving the interval. With these boundary conditions, the discrete mass, $M_f(t)=\sum_i f_i(t)\Delta J_i\simeq\int f(J_z,t)\,dJ_z$, is conserved up to numerical round-off error. At a fixed time $t$, the orbit-averaged action coefficients are obtained by interpolating the precomputed tables for $D_J^{(1)}(J_z,t)$ and $D_{JJ}^{(2)}(J_z,t)$. The numerical flux in equation~\eqref{eq:numerical-flux-fp} is a linear function of the neighbouring cell values of $f$, so we can write the semi-discrete system as
\begin{align}
\frac{d\mathbf f}{dt}
&=
L(t)\mathbf f,
\label{eq:semi-discrete-fp}
\end{align}
where $\mathbf f(t)=\left(f_1(t),f_2(t),\ldots,f_N(t)\right)^T$ is the vector of cell-averaged values. The operator, $L(t)$, acts on $\mathbf{f}(t)$, and is simply the matrix representation of the differential operator $-\partial F_J/\partial J_z$, with
\begin{align}
\left[L(t)\mathbf f\right]_i
&=
-\frac{1}{\Delta J_i}
\left[
F_{i+1/2}(\mathbf f,t)
-
F_{i-1/2}(\mathbf f,t)
\right].
\label{eq:L-operator-definition}
\end{align}
\par
We advance this system using a Crank--Nicolson update \citep{CrankNicolson1947}. For a timestep $\Delta t=t_{n+1}-t_n$, we have
\begin{align}
\left(I-\frac{\Delta t}{2}L_{n+1}\right)\mathbf f_{n+1}&=
\left(I+\frac{\Delta t}{2}L_n\right)\mathbf f_n,
\label{eq:crank-nicolson-fp}
\end{align}
where $L_n=L(t_n)$, and $I$ is the identity matrix. This update treats the drift and diffusion terms time-centrally and is second-order accurate in time for smooth solutions. 
\par
We sample an initial distribution in $(z,v_z)$ to match initial conditions in the test-particle simulation described in \S \ref{sect:test-particle-sim-time}. Each phase-space pair is then mapped to a vertical action using the harmonic approximation. The resulting initial distribution is sharply concentrated at small $J_z$. In this case, a direct Crank--Nicolson step produces small spurious oscillations, because Crank--Nicolson is non-dissipative when applied to poorly resolved or non-smooth initial data. To suppress these transients, we use Rannacher start-up timestepping \citep{Rannacher1984,GilesCarter2006}.
Namely, the first Crank--Nicolson interval is replaced by 4 smaller backward-Euler substeps before switching to the Crank--Nicolson scheme. A backward-Euler substep has the form
\begin{align}
\left(I-\delta t \,L_{m+1}\right)\mathbf f_{m+1}=\mathbf f_m,
\label{eq:backward-euler-substep} 
\end{align}
where $I$ is the identity matrix and $\delta t = \Delta t/4$ is the smaller start-up timestep. Backward Euler is only first-order accurate, but it is strongly damping and therefore smooths the unresolved initial sharp feature. After time $\Delta t$, the evolution is continued with the Crank--Nicolson update in equation~\eqref{eq:crank-nicolson-fp}, retaining second-order accuracy at later times.

\subsection{A time-dependent extension of the equilibrium DF} \label{subsect:time-dependent-df-fit}
\par
The test-particle simulations and the numerical Fokker--Planck solutions numerically provide the full time evolution of the action distribution, but it is useful to describe this evolution with an empirical DF family. We therefore extend the empirical equilibrium DF introduced in Section~\ref{sect: DF-GMC} by allowing its parameters to depend on time. Therefore, the most general form we considered is
\begin{align}
f_z(J_z,t) = A(t)\,
&\exp\left[
-\frac{\alpha(t)}{2}
H_{\gamma(t)}\left(\frac{J_z}{J_c(t)}\right)
\right]
\left(
1+\frac{J_z}{\eta(t)J_{\rm s}(t)}
\right)^{-\eta(t)} \nonumber \\
&\exp\left[
-\left(\frac{J_z}{J_{d}(t)}\right)^{\nu(t)}
\right],
\label{eq:time-dependent-df-full}
\end{align}
where $J_{\rm s}(t)\equiv \sigma_z^2(t)/\Omega_0$, and $H_{\gamma}(x)$ is defined in equation~\eqref{eq:Hgamma}. The normalisation $A(t)$ is fitted independently at each snapshot.
\par
Equation~\eqref{eq:time-dependent-df-full} reduces to the equilibrium DF when the parameters become time-independent and $J_{d}\rightarrow\infty$. The additional final factor is a finite-time cutoff. It is absent in the equilibrium DF, but is required for a single-age population because diffusion has only had a finite time to populate the high-action tail. In the full model, however, both the RLDF-like factor and the cutoff factor can control the high-action fall-off. If both are allowed to vary freely, the optimiser can trade curvature between these two terms, producing nearly indistinguishable fits but qualitatively different temporal parameter trends.
\par
We therefore tested whether the RLDF-like term is required in the time-dependent fits by refitting the snapshots with this term removed. The resulting fits are almost indistinguishable from those obtained with the full model. The fitted trends in $\alpha(t)$, $J_c(t)$, $J_{d}(t)$, and $\nu(t)$ are unchanged, and the residuals remain small, with only a marginal degradation at the youngest snapshots. This shows that the RLDF-like term is not independently constrained in the time-dependent problem. Instead, the cutoff term containing $J_{d}$, records the largest actions populated by GMC scattering over the age of the population, and therefore controls the high-action curvature of the DF. 
\par
We therefore retain the RLDF-like factor for the frozen-in case, where it describes the equilibrium as a function of GMC scaleheight, but omit it from the fiducial time-dependent parametrisation. The finite-time DF then reduces to the product of a thin-GMC-layer cusp and a stretched-exponential envelope. We also fix $\gamma=1$, motivated by the frozen-in equilibrium fits for small cloud scaleheights, $h_c$. With $\gamma=1$, the time-dependent DF becomes
\begin{align}
f_z(J_z,t)
&=
A(t)
\left(
1+\frac{J_z}{J_c(t)}
\right)^{-\alpha(t)/2}
\exp\left[
-\left(\frac{J_z}{J_{d}(t)}\right)^{\nu(t)}
\right].
\label{eq:time-dependent-df}
\end{align}
This reduced DF captures two prominent features of the time-dependent solution, being the low-action cusp produced by the finite GMC scaleheight, and the high-action cutoff produced by incomplete diffusion over a finite time. The pair $\alpha(t)$ and $J_c(t)$ describe the growth of the thin-GMC-layer cusp and its transition scale, while $J_{d}(t)$ and $\nu(t)$ describe the outward motion and shape of the finite-time diffusion envelope.
\par
Finally, we fit only snapshots with $t\geq 2\,{\rm Gyr}$. We avoid early times for fitting our parametrised DF, because they are dominated by the chosen initial conditions. These features are not produced by GMC scattering itself and should not be absorbed into the parameters of the scattering-driven DF. By excluding the first $2\,{\rm Gyr}$, we fit the regime in which the distribution is largely phase-mixed and its evolution is dominated by GMC heating rather than by the initial conditions.

\subsection{Results}
\par
Here, we compare our time-dependent regime and consistency across the test-particle simulation, the numerical solution to the Fokker-Planck equation, as well as the validity of our time-dependent extension of the equilibrium DF. 

\begin{figure}
    \centering
    \includegraphics[width=\linewidth]{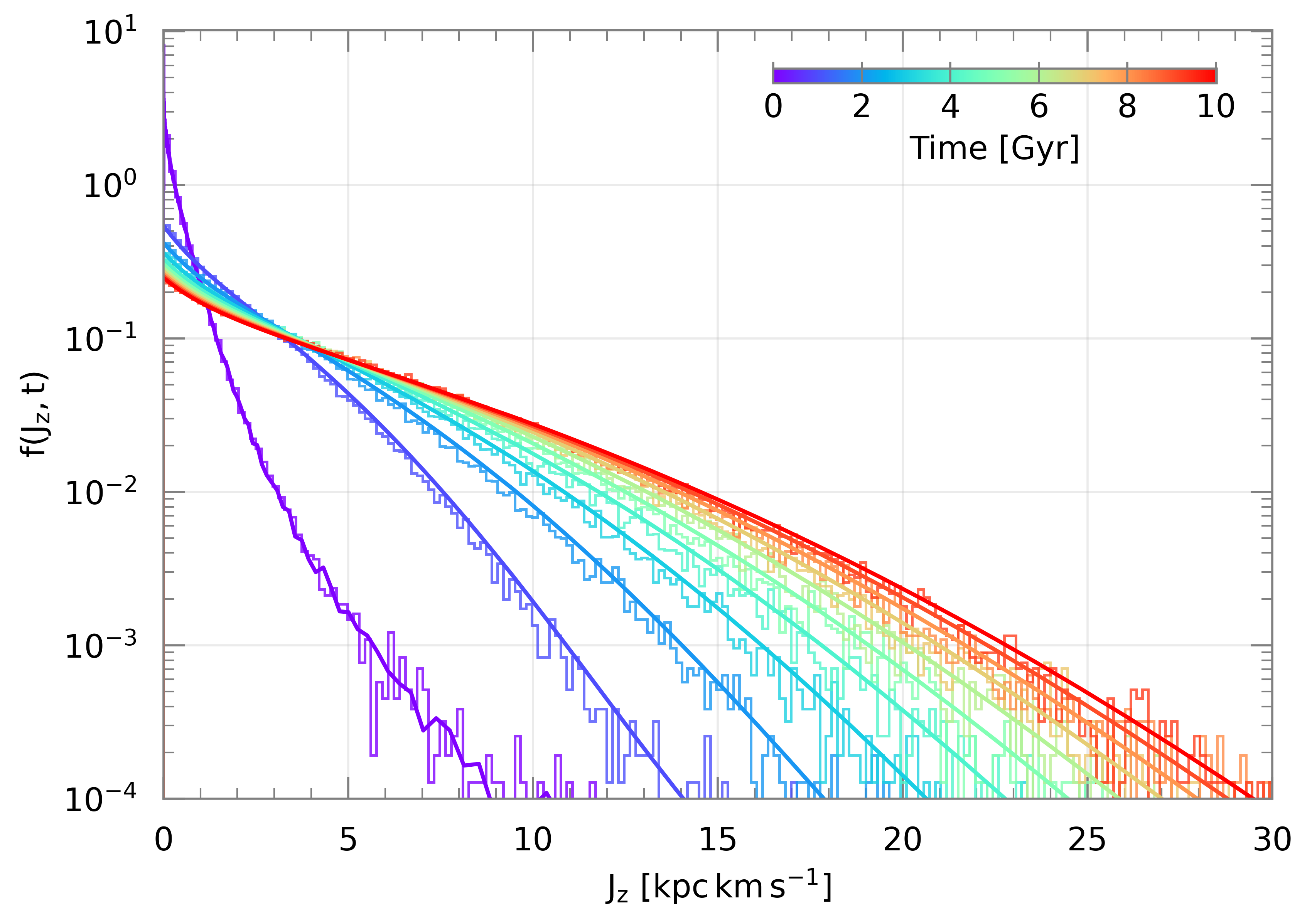}
    \caption{10 Gyr evolution of the action distribution in the harmonic potential, for a coeval population born at time $t=0$, in a time-dependent simulation. Histograms show the test-particle simulation, while solid lines are numerical solutions to the Fokker-Planck equation with time-dependent coefficients and the same initial conditions as the simulation described in \S \ref{sect:test-particle-sim-time}.}
    \label{fig:mono-age-nonstatic-ev}
\end{figure}

Figure \ref{fig:mono-age-nonstatic-ev} displays the time evolution of a coeval population all born at $t=0$, left to evolve in the harmonic potential, with $\Omega_z = 72$ km s$^{-1}$kpc$^{-1}$ as before. Evidently, the distribution at 10 Gyr is far from the steady state, yet it is manifestly non-isothermal. We also run the simulation in the isothermal potential, and find negligible differences in the shape of the DF. We treat the effect of anharmonic potentials in a forthcoming paper. 

\begin{figure}
    \centering
    \includegraphics[width=\linewidth]{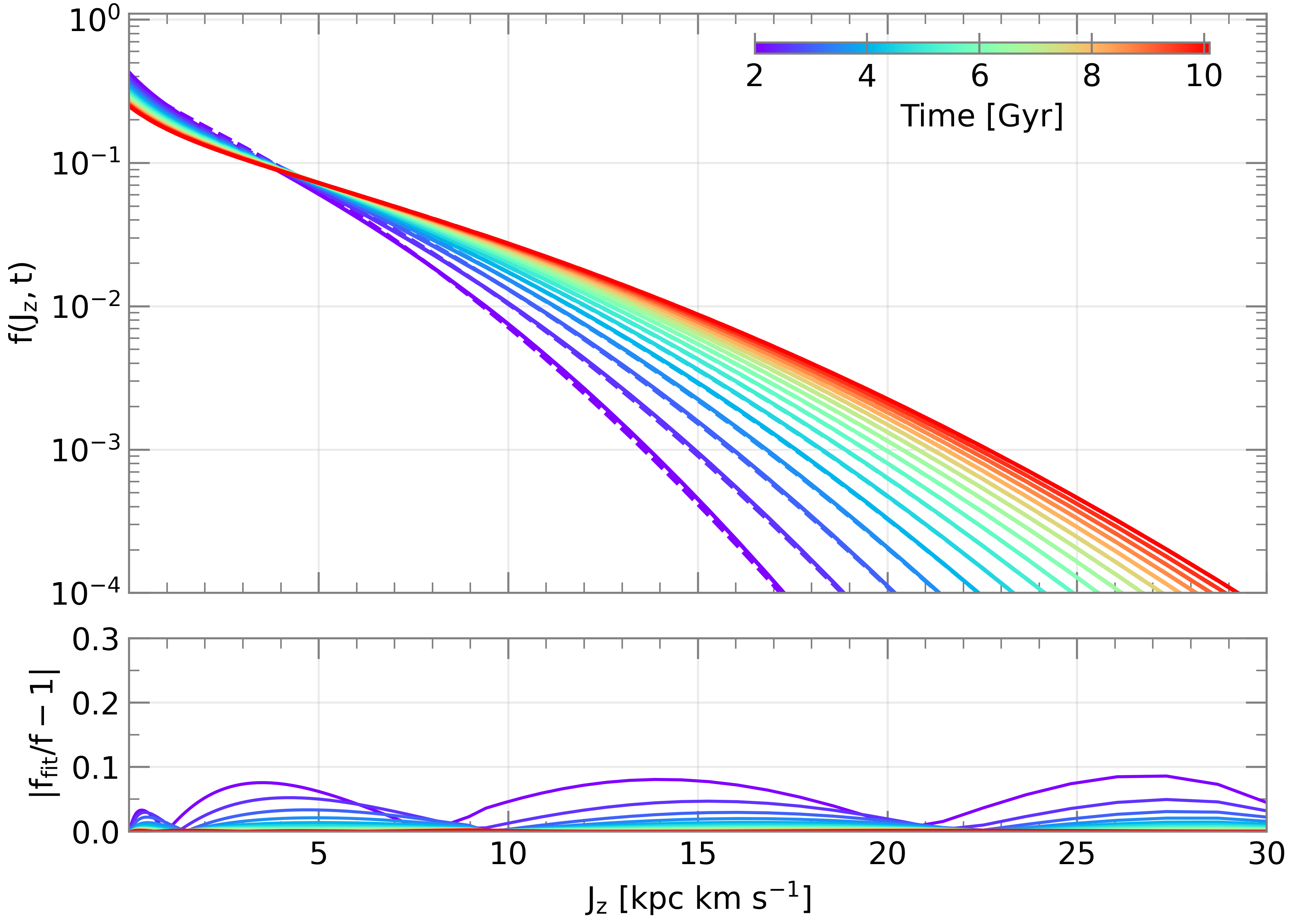}
    \caption{\textit{Top}: Numerical solutions to the Fokker-Planck equation with time-dependent coefficients are shown as solid lines, with the same initial conditions as the simulation described in \S \ref{sect:test-particle-sim-time}, except the initial stellar scaleheights are $h=20$ pc. Dashed lines denote the best fit to the reduced time-dependent DF in equation \eqref{eq:time-dependent-df}. Bottom: Fractional residuals, $|f_{\rm fit}/f-1|$, where $f_{\rm fit}$ is the curve fitted to the analytic DF, and $f$ is the numerical solution to the Fokker-Planck equation. }
    \label{fig:FP_DF_fit}
\end{figure}

\par
To assess the quality of fit to our empirical time-dependent DF in equation \eqref{eq:time-dependent-df}, we solve the Fokker-Planck equation as described in \S \ref{subsec:numerical-fp-J}, except we birth stars with an exponential scaleheight $h=20$ pc, rather than $60$ pc. The colder initial condition is a diagnostic choice. It makes the initial distribution more compact in action, so that the subsequent high-action tail is produced primarily by GMC scattering. If the population is born hotter, the initial DF already contains a broader high-action component, and the solution retains memory of this imposed birth structure for longer. In that case, the fitted DF parameters can absorb features of the initial condition rather than the scattering-driven evolution.  In Figure~\ref{fig:FP_DF_fit}, we plot the numerical Fokker--Planck solutions from $t=2$ to $10\,{\rm Gyr}$ as solid coloured curves, together with the best-fitting empirical DF shown as dashed curves. We fit only snapshots with $t\geq2,{\rm Gyr}$, as justified in \S \ref{subsect:time-dependent-df-fit}. The lower panel shows the fractional residuals, $|f_{\rm fit}/f-1|$, demonstrating that the reduced time-dependent DF captures the numerical solution, although the residuals are largest where the distribution transitions from the initial-condition-dominated core to the scattering-generated high-action tail. However, residuals drop to $\lesssim 0.01$ at later times. We have also verified that replacing the initial exponential vertical profile with a Gaussian sampling in $z$ does not qualitatively change the fitted DF shape at later times, showing that the late-time DF shape is not strongly tied to the precise form of the birth distribution. For evolution with a larger birth exponential scaleheight of $h=60$ pc, the residuals are comparable from $\sim 2.5-3$ Gyr onwards, in light of the above argument.

\begin{figure}
    \centering
    \includegraphics[width=\linewidth]{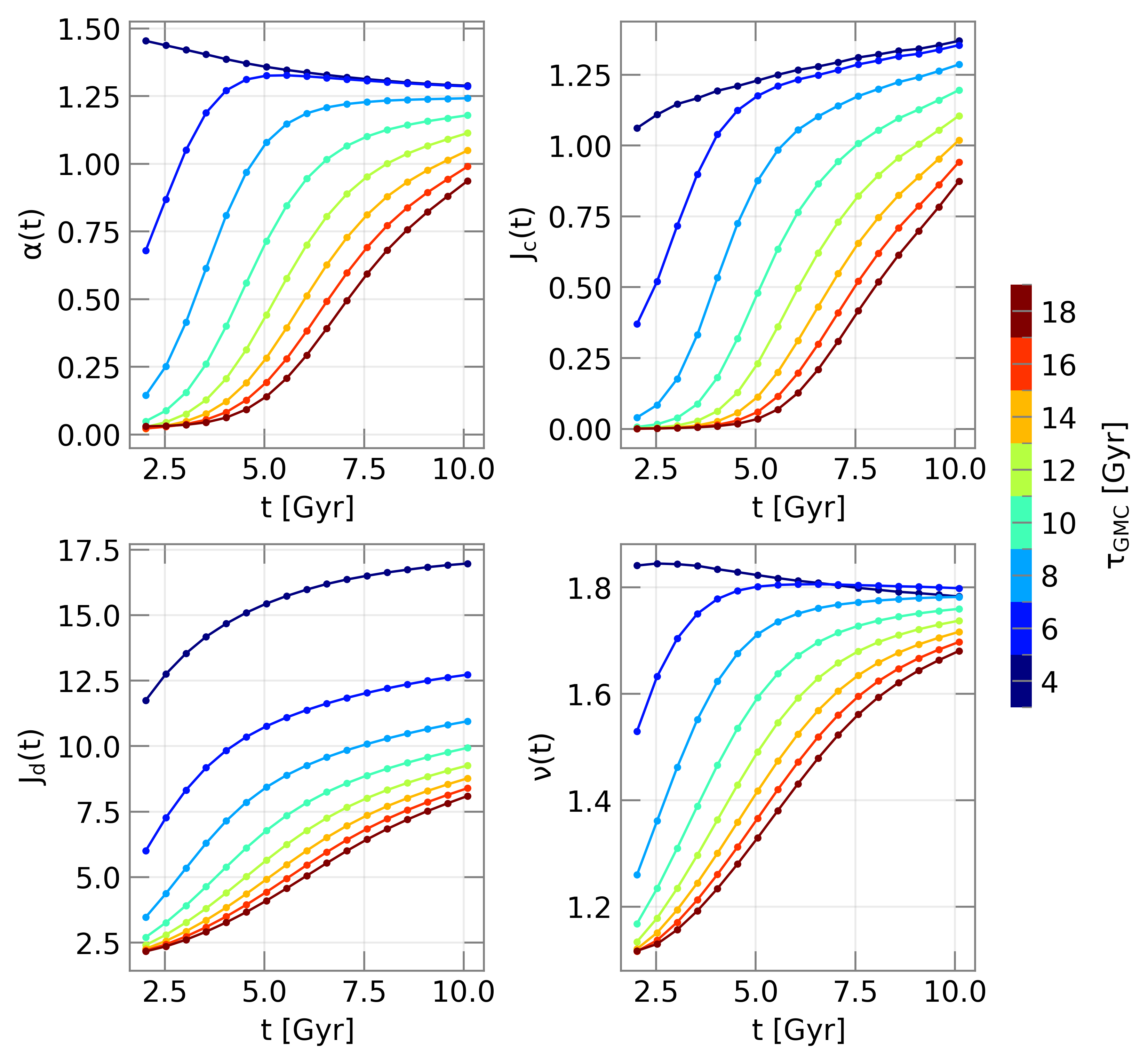}
    \caption{Time evolution of the fitted parameters in the reduced time-dependent DF (equation \eqref{eq:time-dependent-df}) for different GMC decay timescales $\tau_{\rm GMC}$. Colours indicate $\tau_{\rm GMC}$ in Gyr. All models are normalised to the same present-day GMC density at $t\simeq 10,{\rm Gyr}$, so shorter decay times correspond to a much larger GMC density in the past.}
    \label{fig:varying-gmc-params-fit}
\end{figure}

Figure~\ref{fig:varying-gmc-params-fit} shows how the fitted parameters of the reduced time-dependent DF vary with the GMC decay timescale, $\tau_{\rm GMC}$. In these experiments the GMC density histories are anchored to the same present-day value at $t_{\rm present}=10\,{\rm Gyr}$. The decay factor rescales the diffusion coefficients by $\exp[(t_{\rm present}-t)/\tau_{\rm GMC}]$, so smaller $\tau_{\rm GMC}$ corresponds to a much larger GMC density at early times. The parameters, $\alpha(t)$ and $J_c(t)$, grow in time as the cusp develops in the low-action part of the DF. Moreover, the late-time convergence of $\alpha(t)$ and $\nu(t)$ suggests that the shape of the old, scattering-dominated DF is largely controlled by the geometry and velocity dependence of GMC scattering, rather than by the detailed early-time normalisation of the cloud density. Conversely, $J_{d}(t)$ remains more sensitive to $\tau_{\rm GMC}$, suggesting it retains memory of how strongly GMC scattering acted at early times. This behaviour is consistent with the interpretation that the DF evolves towards a self-limited scattering regime, where the shape parameters approach asymptotic values once the population is old and dynamically hot, while the cutoff scale continues to encode the cumulative heating history. A detailed connection between these fitted parameter trajectories, the integrated diffusion exposure, and the self-limiting heating scalings will be developed in a forthcoming paper.

\section{Discussion} \label{sect:discussion}
\par
While diffusion due to scattering of stars by GMCs has been previously investigated, the simplest models of GMC scattering have been shown to be in tension with observational data. Among other early works, \citet{Lacey_1984} found that GMC scattering could not be the primary heating mechanism, and the predicted AVR for the vertical component ($\beta_z \simeq 0.25$), was too low compared to empirically determined values in the Solar neighbourhood ($\beta_z \sim 0.5$; see e.g. \citealt{Aumer_Binney_2009}). The work of Lacey, however, was corroborated by simulations of \citep{Hannien_Flynn_2002}, who recovered a power-law exponent of $\beta_z =0.26$. \citet{Jenkins_Binney_1990} argue it is the combined effect of GMC scattering and scattering off spiral arms that can heat simulations to realistic MW values, although this argument was challenged by later simulations \citep[e.g.][]{ABS_2016} that show that spiral arms provide in-plane heating, but contribute much less overall heating in the vertical direction. More recent chemo-dynamical models found that relinquishing the assumption of a constant heating rate in time, and accounting for increased heating from GMCs at early times, due to higher GMC densities, can resolve these discrepancies \citep[e.g.][]{ABS_2016,Sharma_2021}. 

\subsection{Effect of GMC scattering on the vertical action DF}
\par
In this paper, we address a complementary question: \textit{how does GMC scattering shape the vertical action DF resulting from non-constant diffusion coefficients?} Classical treatments have often adopted simplifications. In particular, \citet{Lacey_1984} neglects cloud motions and approximates relative velocity as simply the stellar velocity. This approximation is best interpreted as a high-velocity asymptotic limit, appropriate when the stellar velocity dispersion is much larger than the cloud velocity dispersion. It is less accurate for young, cold populations, whose dispersions are comparable to those of the clouds. In that regime, neglecting cloud motions overestimates the relative velocity which in turn decreases the first moment (drag) by up to a factor of 2, due to the $V_{\rm rel}^{-3}$ dependence (\textit{see} Appendix \ref{app:sources of error} for more details on sources of error in approximating the moments). Additionally, the second moment can be underestimated by up to $\sim 30 \%$ when neglecting cloud motions. In this regime, gravitational deflections therefore weaken with increasing relative velocity and the heating becomes self-limiting, giving the $\sigma_z\propto t^{1/4}$ behaviour as derived by \citet{Lacey_1984}. However, this limiting behaviour need not describe the full time-evolution from initially cold populations. By accounting for comparable velocity dispersions between young stars and clouds, \citet{Villumsen_1985} recovered a steeper power-law exponent, $\beta_z=0.31\pm0.02$. In our test-particle simulations, we similarly recover an AVR exponent of $\beta_z=0.31$, consistent with \citet{Villumsen_1985}. This suggests that the classical Lacey scaling captures the late-time asymptotic regime, while a more complete treatment is needed to describe how the system approaches that regime.
\par
Moreover, \citet{Binney_Lacey_1988} (\textit{hereafter}, BL88) established the action-space Fokker–Planck framework for secular disc heating and showed that classical cloud scattering does not generally preserve Maxwellian or isothermal distribution functions. Under the assumptions of local, weak scattering by compact clouds and epicyclic stellar orbits, BL88 find that the diffusion tensor in the $(E_R,E_z)$ plane is highly anisotropic. After the velocity ellipsoid approaches an approximately fixed shape, the coeval DF tends towards a late-time self-similar form whose phase-space density falls approximately as a Gaussian in the energies. In a harmonic vertical potential, $E_z=\Omega_zJ_z$, which corresponds to a Gaussian fall-off in vertical action, rather than to the pseudo-isothermal form $f_z\propto\exp(-J_z/J_0)$. Our results are qualitatively consistent with the non-isothermality of BL88, but show that the vertical DF is \textit{not} simply Gaussian in action. The high-action tail of the GMC-scattered DF is depleted relative to a pseudo-isothermal exponential, and our fitted cutoff exponent, $\nu\simeq1.7$ at late times, is slightly lower than the Gaussian expectation of $\nu=2$ (cf. equation \eqref{eq:time-dependent-df}). However, the finite thickness of the GMC layer also imprints a low-action cusp, which redistributes probability toward small actions and changes the normalisation and curvature of the outer DF, so that the resulting distribution cannot be described as a pure Gaussian cutoff in $J_z$. BL88 therefore provide an important precedent for non-isothermal coeval DFs produced by cloud scattering. Here we build on that result by computing the vertical action coefficients directly and accounting for the confined GMC layer.
\par
BL88 distinguish between two physically different drift--diffusion relations in action space. For an externally imposed stochastic perturbing potential, the first-order coefficient follows from the inhomogeneity of the diffusion tensor itself, $A_i=(1/2)\partial A^2_{ij}/\partial J_j$, giving the conservative diffusion equation (their equations 3.10--3.11). This relation does not assume, or select, a Gibbs equilibrium. Conversely, when the scatterers recoil and are assumed to have a fixed thermal equilibrium distribution, detailed balance gives the fluctuation--dissipation relation of equation (3.12), which contains an additional friction term proportional to $\beta A^2_{ij}\Omega_j$. The zero-flux solution of this relation is $f\propto \exp(-\beta H)$. In the harmonic approximation, this yields exactly the pseudo-isothermal form $f_z\propto\exp(-J_z/J_0)$. Thus, caution is required when using the fluctuation--dissipation relation as a closure for GMC scattering, because it effectively pre-selects an isothermal equilibrium, rather than deriving the stationary DF from the true action dependence and geometry of the GMC scattering coefficients.
\par
A more recent test-particle simulation of GMC was investigated in \citet{Tremaine_2023}, although their aim was to study the survival of the \textit{Gaia} snail \citet{Antoja_2018}, rather than to derive the detailed DF generated by GMC scattering. In their model, the small-scale perturbations are represented by zero-mean Gaussian kicks in the canonical vertical variables, with a constant diffusion amplitude independent of phase-space coordinates. Since $J_z=(q^2+p^2)/2$, this stochastic process implies a pseudo-isothermal vertical DF, irrespective of the strength of scattering, or the spatial GMC distribution. Such models may be adequate for exploring generic stochastic heating or the erasure of phase-space spirals, but it should not be interpreted as evidence that realistic GMC scattering generically produces an isothermal vertical DF. This distributed and velocity-independent scattering implies a process very different from more realistic GMC scattering, as we have shown. 
\par
In this work the first- and second-order velocity moments are derived from two-body star--cloud scattering. The resulting coefficients are strongly non-linear in $v_z$ and become age-dependent through the evolving in-plane stellar dispersions and the declining GMC density. After orbit averaging, this produces non-linear action-space coefficients and therefore non-isothermal DFs. The frozen-coefficient zero-flux solutions develop a low-action cusp set by the finite thickness of the GMC layer and non-exponential behaviour at larger actions. In the fully time-dependent problem, we found a more accurate, yet simple, parameterisation via an empirical DF, given in equation \eqref{eq:time-dependent-df}, which was also revealed to be non-isothermal.
\par
Our work reveals an important consequence of the shape of the DF. Even if a simplified stochastic model reproduces a similar AVR or heating efficiency, it may predict the wrong response amplitude for the disc if it assumes the wrong equilibrium DF. This is especially relevant in problems where the response depends sensitively on the detailed action-space structure of the DF, rather than only on its velocity dispersion. For example, non-linear or growing perturbations can be carried disproportionately by stars at small actions, where our DF differs most strongly from a pseudo-isothermal form. In such cases, initialising the disc with an isothermal action DF may misestimate the susceptibility of the disc, and therefore bias the inferred strength or evolution of additional perturbations.

\subsection{Effect of a constant Coulomb logarithm}
\begin{figure}
    \centering
    \includegraphics[width=\linewidth]{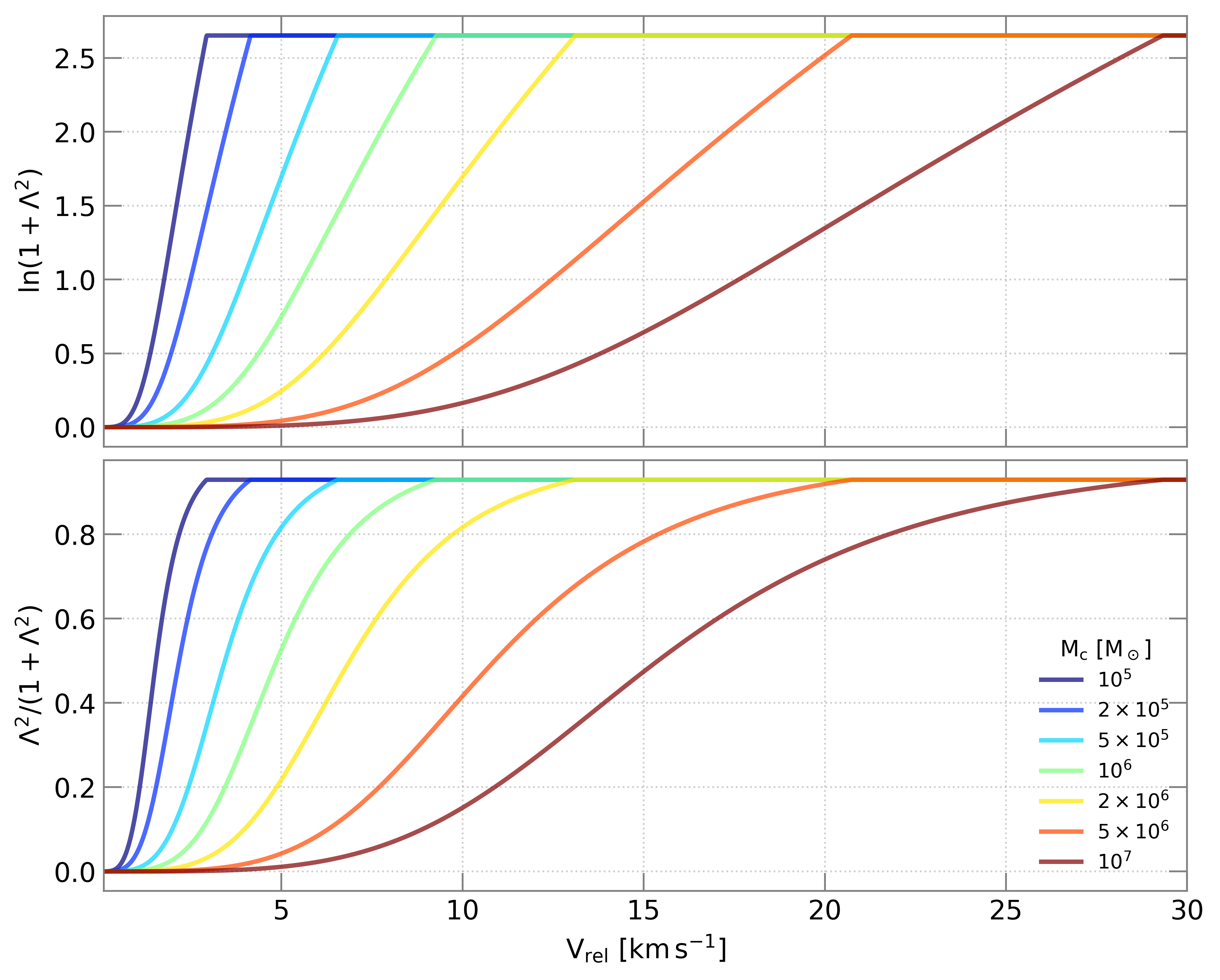}
    \caption{$\ln (1+\Lambda^2)$ (\textit{top}) and $\Lambda^2 / (1+ \Lambda^2)$ (\textit{bottom}) plotted as a function of relative velocity magnitude between stars and clouds, $V_{\rm rel}$, for a range of different cloud masses, $M_c$, labelled in the legend.}
    \label{fig:lambda-vs-vrel}
\end{figure}
\par
The Coulomb logarithm, $\ln \Lambda$, appears implicitly in the Fokker-Planck equation through the velocity moments. While not adequately constrained in the literature, a common choice of $\Lambda$ is typically a \textit{large} value, such as the size of the system, which is then taken out as a constant of the integrand in the calculation of velocity moments \citep[e.g.][]{RMJ_1957,Henon_1973}. \citet{Merritt_2001} investigates where this approximation fails, in the context of dynamical friction, and discussions on the Coulomb logarithm followed in \citet{Hashimoto_2003}, \citet{Just_2005}, \citet{Petts_2016}, although to our knowledge, an explicit treatment of the velocity-dependent Coulomb logarithm in the context of GMC scattering in galactic discs has not yet been investigated. For dynamical friction, in a classical derivation, one assumes an infinite homogeneous background of field stars; in a disc with a thin GMC layer, these assumptions do not hold and thus a confined $b_{\max}$ better captures the geometry of the problem. 
\par
In this paper, we do not assume a constant $\Lambda$, and we adopt improved prescriptions of $b_{\rm max}$, and hence, $\Lambda$, as a function of relative velocity between stars and clouds, which is plotted in Figure \ref{fig:lambda-vs-vrel}. We plot $\ln (1+\Lambda^2)$, which appears in the integrand of the first moment of velocity, and $\Lambda^2 / (1+ \Lambda^2)$, which appears in the integrand of the second moment. Our new propositions for $b_{\rm min}$ and $b_{\rm max}$ are laid out in equations (\ref{eq:bmin}) and (\ref{eq:bmax}), respectively. At small relative velocities, $b_{90}=GM_c/V_{\rm rel}^2$ becomes large, and can exceed $b_{\rm max}$, implying, $\Lambda<1$. This should not be interpreted as a negative or ordinary Coulomb logarithm. Rather, it means that there is no resolved interval of impact parameters satisfying $b_{\rm min} < b < b_{\rm max}$ over which the encounter can be treated as a weak, local, independent scattering event. In this regime, we suppress the weak-scattering contribution. Physically, slow encounters are either softened by the finite cloud size, fail the impulse approximation, or behave as part of the mean-field background potential rather than as random two-body kicks.
Nonetheless, for small relative velocities, where $\Lambda > 1$, or $\ln (1+\Lambda^2) \gtrsim 0.69$, Figure \ref{fig:lambda-vs-vrel} reveals the terms entering the velocity moments are strongly non constant, with effects most pronounced for heavier clouds, and thus, approximating $\Lambda$ by a large constant erases this behaviour. Nearly co-moving encounters are then weighted almost as strongly as high–$V_{\rm rel}$ encounters, which overestimates the contribution of slow encounters. This primarily biases the low-action part of the orbit-averaged coefficients, because low-$J_z$ stars have small vertical velocities and remain embedded in the GMC layer. Since the zero-flux DF depends on both $1/\overline{D^{(2)}_{JJ}}$ and the ratio $2\overline{D^{(1)}_J}/\overline{D^{(2)}_{JJ}}$, this changes the inner logarithmic slope of $f_z(J_z)$. A constant Coulomb logarithm would therefore tend to over-diffuse cold orbits, weaken or otherwise distort the GMC-layer cusp, and drive the inferred DF closer to a pseudo-isothermal form than is produced in this work.

\par
In our prescription, $b_{\rm max}$ is tied to the vertical extent of the GMC layer, because encounters on scales much larger than the cloud scaleheight are no longer well described as local, independent two-body kicks. Instead, they contribute to the smooth or slowly varying GMC mean field, whose effect on the vertical action is adiabatically suppressed. Thus, contrary to the classical homogeneous-background treatment, $b_{\rm max}$ should not be taken to be a large system-size scale in the disc problem. It should rather be of order the geometric thickness of the cloud layer, or the mean separation of clouds.
\par
The large-$h_c$ distributions presented in this paper are therefore useful only as an artificial limiting experiment. As $h_c$ is increased, the model admits progressively more distant encounters, the scattering becomes more nearly randomised along the orbit, and the resulting DF approaches the pseudo-isothermal limit. This behaviour demonstrates the sensitivity of the equilibrium DF to the assumed choice of $b_{\rm max}$, but it should not be interpreted as a realistic equilibrium for the Milky Way. For the realistic GMC populations, where $h_c\simeq 50$--$100,{\rm pc}$, we propose the use of a small, geometrically motivated $b_{\rm max}$.

\subsection{Limitations}
\par
The most obvious limitation of our work is undoubtedly the restriction to one dimension. In particular, we neglect explicit coupling between vertical and in-plane actions, $J_R$ and $J_\phi$, nor do we include resonant effects (e.g. from spiral arms or the Galactic bar). In a three-dimensional treatment of the Fokker-Planck equation, the diffusion tensor will contain cross-terms, allowing for an exchange between vertical and in-plane actions, leading to a moderate dependence of scattering terms on
$J_R$ and $J_\phi$. By construction of our model, we neglect this coupling, and evolve $J_z$ independent of in-plane actions, in a fixed potential, $\Phi(z)$. We implement GMC scattering only through the vertical components of the drift and diffusion coefficients. 
\par
As a simple thought experiment, consider stars and clouds that both remain close to the Galactic midplane. Then, if the relative velocity (magnitude) is high, the star will scatter considerably in the plane, but the mean vertical force is symmetric about the Galactic midplane, so the \textit{first} vertical moment $D^{(1)}_{z}$ is close to zero. There is, however, a non-zero \textit{second} moment, $D^{(2)}_{zz}$ which would cause diffusion in $v_z$, and thus vertical heating. A one-dimensional model can include this effect only after averaging over the assumed in-plane velocity distribution; it cannot follow the correlated exchange of energy between $J_z$, $J_R$, and $J_\phi$ in individual encounters.
\par
Another limitation is the Fokker--Planck formulation itself, which assumes that the evolution is produced by many weak encounters. This approximation becomes questionable for slow relative encounters, where $b_{90}=GM_c/V_{\rm rel}^2$, becomes comparable to or larger than $b_{\max}$. Equivalently, encounters cease to be weak for $V_{\rm rel}\lesssim (GM_c/b_{\max})^{1/2}$. For $M_c=10^6\,{\rm M_\odot}$ and $b_{\max}=100\,{\rm pc}$, this gives $V_{\rm rel}\simeq6.6\,{\rm km\,s^{-1}}$, comparable to the velocity dispersion of young stars and clouds, and therefore to the low-action region where the GMC-layer cusp forms. Thus, in this regime, a direct treatment of individual encounters would be more appropriate than a purely diffusive approximation.
\par
These simplifications allow us, however, to isolate how the \textit{vertical} diffusion coefficients, including their non–linear velocity dependence, shape the steady–state action distribution, without the additional complexity of other MW components or radial migration. We expect these to moderately affect the DF shape, but expect it not to alter our conclusions on the non-constancy of $\Lambda$ and non-isothermality of the asymptotic DF. A fully three-dimensional treatment is left for future work.

\section{Conclusion} \label{sect:conc}
\par
We have revisited the commonly adopted pseudo-isothermal distribution for vertical motions in the stellar disc, which closely relates to the Maxwellian/Schwarzschild, velocity distribution and implies a thermal equilibrium. GMCs reside in a narrow layer around the midplane and stars are far from equipartition with the GMCs, so the scattering is neither spatially uniform nor are the scatterers in thermal equilibrium. This motivated us to investigate whether GMC-driven vertical heating should produce an exponential DF in $J_z$ at all.
\par
In this paper, we found that the common assumption of a constant Coulomb logarithm $\Lambda$ should be dropped, as it is expected to change by an order of magnitude, with most pronounced non-constancy at small relative velocities, between stars and clouds. This contributes to the non-isothermality brought on by non-linear drag and non-constant diffusion terms in velocity. Further, the narrow height of GMCs drives spatially dependent heating rates, concentrating the action distribution towards the Galactic midplane.
\par
We first considered an idealised razor-thin GMC layer. A simple statistical argument then showed that the usual pseudo-isothermal form admits an additional low-action factor, $J_z^{-\alpha}$, giving a DF steeper than an exponential DF at small $J_z$. This simple model already demonstrates that concentrating scattering near the midplane naturally produces non-isothermal structure in action space.
\par
We also revised the velocity moments from the mechanics of two-body scattering and simulated the star-cloud interactions in a more realistic extended GMC distribution using a diffusive Fokker-Planck regime. We compute stationary solutions, and extend to fully time-dependent models, including a declining GMC density and the growth of in-plane stellar dispersions. We find vertical action distributions are well described by the reduced time-dependent DF in equation~\eqref{eq:time-dependent-df}. We propose this form as a practical alternative to a pseudo-isothermal DF for GMC-scattered coeval populations. For the oldest populations in our fiducial MW-like models, the shape parameters become slowly varying, with representative late-time values $\alpha_{\rm old}\sim 1$ and $\nu_{\rm old}\simeq 1.7$--$1.8$, while $J_{c,\rm old}$ should be taken from the late-time plateau of the fitted models. The cutoff scale $J_{d}$ should not be treated as universal, since it is most sensitive to the accumulated scattering history, the GMC density evolution, and the cloud mass spectrum. In applications, it should be calibrated either to a chosen scattering model or to the observed AVR.

\vspace{-2.0em}
\section*{Supplementary Material}
Supplementary material is available at MNRAS online. 
\vspace{-2.0em}
\section*{Acknowledgments}
We thank Walter Dehnen for insightful conversations. MD acknowledges support from a UCL ORS/GRS PhD scholarship. RS acknowledges the generous support of a Royal Society University Research Fellowship.
\vspace{-2.0em}
\section*{Data availability}
The code used to produce the results are available from the corresponding author upon request.
\vspace{-2.0em}

\bibliographystyle{mnras}
\bibliography{example} 



\appendix
\counterwithin{figure}{section}

\section{Ensemble relative speed} \label{app: relative speed}
Here, we evaluate the ensemble relative speed integral in equation (\ref{eq:v_rel_integral}) using a Monte Carlo approach, and compare to our approximation in equation (\ref{eq:v_rms}). Again, since our simulation is one dimensional, we need to draw in-plane velocities from a distribution $\mathbf{g}(\mathbf{v_\perp})$. To do this, we take $R_\odot = 8.27$ kpc as we are interested in the Solar neighbourhood, and the speed of the LSR as $\Theta_0 = 238$ km s$^{-1}$ \citep{Schonrich_2012}. As in \citet{Sharma_2021}, we model the velocity dispersion (in $R$ and $z$) as a function of angular momentum, $L_z$, and age $\tau$, ignoring metallicity dependence. We also neglect the weak dependence on vertical height $z$, since our scattering is concentrated near the midplane. We adopt equations (4) and (5) from \citet{Sharma_2021}, assuming a power-law dependence on age, and adopt their best fit parameters. We adopt a circular velocity curve $v_c(R,z)$ as in \citet{Sharma_2014} and use their best fit parameters. Finally, we use the asymmetric drift equation \citep[see][equation 4.227]{BT2008} to obtain the mean azimuthal motion and adopt $\sigma_\phi \approx 0.59 \sigma_R$ \citep{BT2008}. We then sample $v_R$ and $v_\phi$ consistent with these dispersions, to generate a  representative mock sample of in-plane velocities, $\mathbf{g(v}_{\perp}| \tau)$, for ages $\tau \in [0,10]$ Gyr. The age-velocity dispersions are fit to a power law \citep{Aumer_Binney_2009},
\begin{align}
    \sigma_i(\tau) = \sigma_{0,i} \left( \frac{\tau + 0.1 }{10 \text{ Gyr} + 0.1} \right)^{\beta_i},
\end{align}
where $i = R,\phi$, $\sigma_{0,i}$ is the present-day velocity dispersion, and $\beta_i$ is the power-law exponent. We find best fit parameters, $\sigma_{0,R} = 40.68$ km s$^{-1}$, $\beta_R = 0.244$,
$\sigma_{0,\phi} = 23.99$ km s$^{-1}$, and $\beta_\phi = 0.244$, consistent with observational constraints in the Solar neighbourhood \citep{Sharma_2021}.
\par
\begin{figure}
    \centering
    \includegraphics[width=\linewidth]{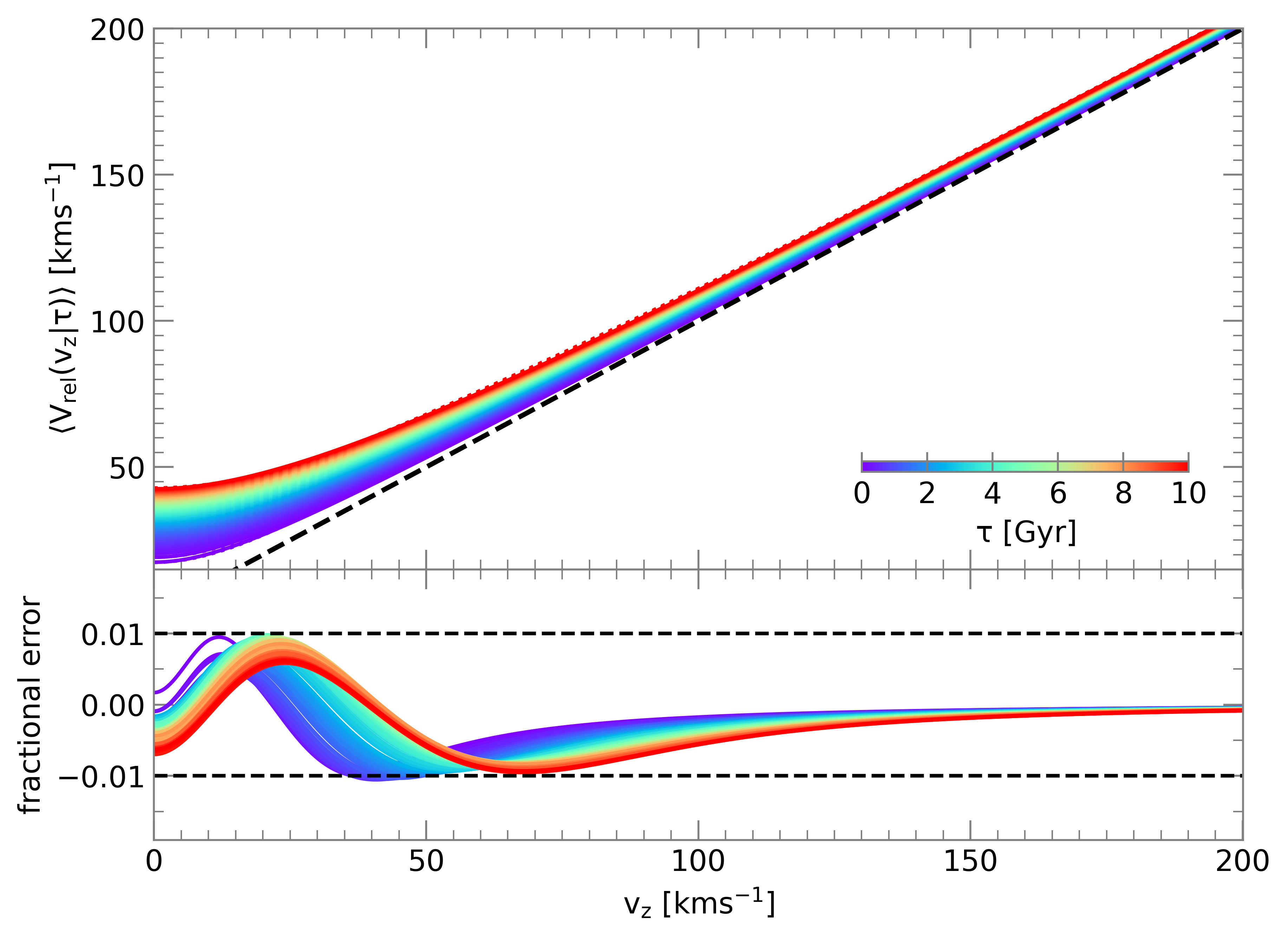}
    \caption{\textit{Top}: Solid lines are exact ensemble averages of relative velocity as a function of $v_z$ (equation (\ref{eq:v_rel_integral})), and dashed lines are the corrected RMS approximation, coloured by different ages. The black dashed line is $y=x$ line. \textit{Bottom}: Fractional error between the exact integral and the corrected RMS approximation. }
    \label{fig:vrel-vs-vz-age}
\end{figure}

Now, we adopt a Monte Carlo approach to estimate the integral in equation (\ref{eq:v_rel_integral}). Once we obtain a mock sample of in-plane velocities, we draw components of the cloud velocity, assumed to be isotropic. Hence, each component is sampled from a Gaussian centred at $0$, with dispersion $\sigma = 6$ km s$^{-1}$. Then, for each fixed $v_z$ and $\tau$, we iterate $N=10^6$ times to compute $V_{\rm rel} = |\mathbf{V}_c - \mathbf{v_\star}|$, and then finally compute the mean over all draws. We repeat this process for velocities $v_z \in [0, 200]$ km s$^{-1}$, with a spacing of $\Delta v_z = 0.1$ km s$^{-1}$. The top panel of Figure \ref{fig:vrel-vs-vz-age} (solid lines) display the ensemble relative velocities as a function of vertical velocity, $v_z$, coloured by different ages. The black dashed line denotes the $y=x$ line, implying the ensemble relative velocity becomes dominated by $v_z$ at large speeds. 
\par
We propose the use of an RMS approximation in equation (\ref{eq:v_rms}), given that velocity components of both stars and the clouds are approximately Gaussian. For $\sigma_R(\tau)$ and $\sigma_\phi(\tau)$ we use the velocity dispersions computed from our Monte Carlo sampling at each age.

We first adopted an approximation,
\begin{align}
    V_{\rm rms} = \sqrt{v_z^2 + 3\,\sigma^2 + \sigma_R^2(\tau) + \sigma_{\phi}^2(\tau)},
\end{align}
although, unsurprisingly, the RMS approximation overestimated relative velocities for small $v_z$ by up to $\sim 10\%$. This incentivised adding a dimensionless correction factor $C_V$, which we empirically chose to have the form,
\begin{align} \label{eq:C1}
    C_V(v_z, \tau) = 1 - a_0 \left( 1 -a_1 e^{-\tau / \tau_{\rm 1}} \right) 
\exp\Bigg[-\frac{v_z^2}{\sigma(\tau)^2}\Bigg], 
\end{align}
where $\sigma(\tau) = \sigma_0 \left[ 1 - \sigma_1 \, e^{-\tau / \tau_{\rm 2}} \right]$. This form best suppressed errors at small $v_z$ and the weak age-dependence of the error. The best fit free parameters are given in Table \ref{tab:c1}. The dashed lines in the top panel of Figure \ref{fig:vrel-vs-vz-age} plot the corrected RMS approximation, with fractional error between the exact MC evaluation and the approximation shown in the bottom panel of Figure \ref{fig:vrel-vs-vz-age}. All errors are within $\sim 1 \%$, and thus we justify the use of the approximation in equation (\ref{eq:v_rms}) for computational efficiency, and find negligible effects on the resulting distribution of stars. 

\begin{table}
\centering
\caption{Best-fit parameters for correction $C_V(v_z, \tau)$ (equation (\ref{eq:C1})).}
\label{tab:c1}
\begin{tabular}{lcc}
\hline
Parameter & Value & Units \\
\hline
$a_0$        & 0.1144  & ---        \\ 
$a_1$        & 0.1242  & ---        \\ 
$\sigma_0$   & 28.8507 & km/s      \\ 
$\sigma_1$   & 0.4909  & ---       \\ 
$\tau_1$     &  $1.0039$   & Gyr       \\ 
$\tau_{2}$  &   3.8811     & Gyr   \\ 
\hline
\end{tabular}
\end{table}

\section{Derivation of velocity moments} \label{app:moments-derivation}
\par
Here, we derive the moments of velocity change from the mechanics of two-body scattering (\textit{see} \citet{Henon_1973} for a similar derivation). We begin with all of the assumptions outlined in \S \ref{subsect:moments of velocity}. Now, consider an incident star of mass, $m_1$ approaching a GMC of mass $M_c$, with impact parameter, $b$. We work in the instantaneous rest frame of the GMC, in which the incoming star approaches with relative velocity  $\mathbf{V}_{\rm rel}=\mathbf{V}_\star-\mathbf{V}_c$, and $\mathbf{V}_{\star}$ and $\mathbf{V}_{c}$ are the star and cloud velocities, respectively. Denote $V_{\rm rel} = |\mathbf{V}_{\rm rel}|$ as the magnitude of the relative velocity. The initial true anomaly, $\phi_0$, is measured as the angle between the initial velocity and the pericenter approach of the deflection. Upon passing on a hyperbolic orbit, the star changes direction, but its magnitude is unchanged. The components of velocity parallel to the initial motion, $\Delta V_{||}$, and perpendicular, $\Delta V_{\perp}$, are simply given by,
\begin{align}
    \Delta V_\perp &= V_{\rm rel} \sin \theta_d \label{eq:v_perp} \\
    \Delta V_{||} &= -(V_{\rm rel}-V_{\rm rel} \cos \theta_d) =  -V_{\rm rel} (1-\cos \theta_d). \label{eq:v_par}
\end{align}

Observe that to now acquire a perpendicular velocity component towards the GMC, then it must slow down the star in the parallel direction, hence the minus sign. To eliminate the $\theta_d$ dependence, we can determine the true anomaly in terms of other variables. The standard orbit equation for conic sections is given by,
\begin{align}
    r = \frac{l}{1+e\cos \phi},
\end{align}
where $l$ is the \textit{semi-latus rectum} and $e$ is the \textit{eccentricity}. We can write $l = h^2 / (G(M_c +m_1))$, where $h = bV_{\rm rel}$ is the angular momentum per unit mass. It can be shown that the true anomaly for a hyperbolic deflection in the limit $r \rightarrow \infty$, is given by,
\begin{align}
    \tan \phi_0 = \frac{bV_{\rm rel}^2}{G(M_c+m_1)}.
\end{align}
Using the relation, $\theta_d = \pi-2\phi_0$, and substituting into equations (\ref{eq:v_perp}) and (\ref{eq:v_par}), we have,
\begin{align}
    \Delta V_\perp^{\rm (COM)} &=  \frac{2bV_{\rm rel}^3}{G(M_c+m_1)} \, e^{-2}\label{eq:v_perp_new} \\
    \Delta V_{||}^{\rm (COM)} &= -2V_{\rm rel} \,e^{-2} \label{eq:v_par_new},
\end{align}
where,
\begin{align}
    e^2 = 1 + \frac{b^2V_{\rm rel}^4}{G^2(M_c+m_1)^2}.
\end{align}
If we define,
\begin{align} \label{eq:b90}
    b_{\rm 90} \equiv \frac{G(M_c+m_1)}{V_{\rm rel}^2},
\end{align}
corresponding to a maximum deflection of $\theta_d = 90^\circ$, then we can write the eccentricity as,
\begin{align}
    e^2 = 1 + \frac{b^2}{b_{\rm 90}^2}.
\end{align}
Finally, the change of the relative velocity is converted into the change of the stellar velocity by multiplying by $M_c/(M_c+m_1)$.
Perpendicular and parallel to its initial motion, we have,
\begin{align}
    \Delta v_{m,\perp} &=  \frac{2b M_cV_{\rm rel}^3}{G(M_c+m_1)^2} \, e^{-2}\label{eq:v_perp_star} \\
    \Delta v_{m,||} &= -\frac{2M_cV_{\rm rel}}{M_c+m_1} \,e^{-2} \label{eq:v_par_star}.
\end{align}
In the weak-scattering limit, $b\gg b_{90}$, we have $e^{-2}\simeq b_{90}^2/b^2\propto V_{\rm rel}^{-4}$. Hence $\Delta v_{m,\perp}\propto V_{\rm rel}^{-1}$ while $\Delta v_{m,\parallel}\propto V_{\rm rel}^{-3}$. These scalings should not be extrapolated to the strong-encounter or low-$V_{\rm rel}$ regime, where $b\lesssim b_{90}$ and the weak-scattering approximation breaks down.
\par
Now, we compute moments of velocity change. Firstly, consider a star passing a sea of clouds moving at fixed relative speed, $V_{\rm rel}$. In a time interval $\Delta t$, the average rate of an encounter with a GMC with impact parameters in the range, $b,b+\text{d}b$ is,
\begin{equation}
    \text{d}p = 2\pi \,  n_c \, V_{\rm rel} \, \Delta t \, b \, \text{d}b ,
\end{equation}
where $n_c$ is the local (spatial) number density of GMCs at the position of the star, given in equation (\ref{eq:cloud-density}). Then, the first moment of velocity change, for a fixed $V_{\rm rel}$ is computed as,
\begin{align} \label{eq: first-moment-general}
    \langle \Delta v \rangle_{V_{\rm rel}} = \int \text{d}b \, \Delta v \, 2\pi \,  n_c \, V_{\rm rel} \, \Delta t \, b
\end{align}
with a similar expression for the second-order moment. For the analytic point-mass derivation below, the impact-parameter integrals are first written over the range $0<b<b_{\rm max}$, which gives the usual Coulomb factors in terms of $\Lambda_{90}=b_{\rm max}/b_{90}$. In the simulations, however, we use a regulated weak-scattering prescription with a finite lower cutoff, $b_{\rm min}=\max(r_c,b_{90})$ and define $\Lambda=b_{\rm max}/b_{\rm min}$. This excludes unresolved close passages through the cloud, and removes the regime in which there is no available interval of weak, local encounters. The modification mainly changes the normalisation and low-$V_{\rm rel}$ behaviour of the velocity moments, but it does not change the qualitative result that the drift and diffusion coefficients are strongly velocity-dependent.
\par
In particular, we are interested in the $z-$component of the velocity change. Denote $\hat{\mathbf{e}}_{||} = \mathbf{V}_{\rm rel} / V_{\rm rel}$ as the direction of the initial motion. Then, $\mu = \hat{\mathbf{e}}_{||} \cdot \hat{\mathbf{z}} =(v_z - V_{\rm c,z}) / V_{\rm rel}$ is the projection of the star's vertical speed in the $z-$direction, and $v_z$ and $V_{\rm c,z}$ are the $z-$components of the star's and cloud's velocity, respectively. Using equations (\ref{eq:v_perp_star}) and (\ref{eq:v_par_star}), the $z-$component of velocity change is written as,
\begin{align}
    \Delta v_z = \mu \Delta v_{m,||} + (\hat{\mathbf{n}}\cdot \hat{\mathbf{z}}) \Delta v_{m,\perp},
\end{align}
where $\hat{\mathbf{n}}$ is perpendicular to $\hat{\mathbf{e}}_{||}$ and is randomly oriented. Under the approximation $M_c \gg m_1$, we have $(M_c + m_1)^2 \approx M_c^2$. Taking the first moment as per equation (\ref{eq: first-moment-general}), the perpendicular component cancels out due to symmetry, and we are left with,
\begin{equation}
    \langle \Delta v_z \rangle_{V_{\rm rel}} = -2\pi G^2 M_c^2 n_c \Delta t  \ln \left [ 1 +\Lambda(V_{\rm rel})^2 \right] \frac{v_z - V_{\rm c,z}}{V_{\rm rel}^3},
\end{equation}
where $\Lambda(V_{\rm rel}) \equiv b_{\rm max}/b_{\rm 90}(V_{\rm rel})$, and $b_{\rm 90}(V_{\rm rel})$ is defined in equation (\ref{eq:b90}). This is conditional on a fixed $V_{\rm rel}$. Now, consider a distribution of GMCs whose velocities, $\mathbf{V}_{\rm c}$, are isotropic. We write the three-dimensional velocity distribution as 
\begin{align}
    F_c(\mathbf V_{\rm c}) = \frac{1}{(2\pi\sigma^2)^{3/2}}
\exp\left(-\frac{|\mathbf V_{\rm c}|^2}{2\sigma^2}\right),
\end{align}
The ensemble-averaged first moment is then
\begin{align} \label{eq: dv-exact}
     \langle \Delta v_z \rangle = &-2\pi G^2 M_c^2 n_c \Delta t \, \times \nonumber \\
     &\int \mathrm d^3\mathbf V_{\rm c}\, F_c(\mathbf V_{\rm c})\frac{v_z - V_{\rm c,z}}{V_{\rm rel}^3} \, \ln \left[ 1 + \Lambda(V_{\rm rel})^2 \right],
\end{align}
$V_{\rm rel} = |\mathbf{V}_\star-\mathbf{V}_{c}|$ as before.
\par
We proceed similarly for the second moment. The $z-$component is written as,
\begin{align}
    (\Delta v_z)^2 = \mu^2 \Delta v_{m,||}^2 +  \left(\frac{1-\mu^2}{2}\right)  \Delta v_{m,\perp}^2
\end{align}
and integrating over all impact parameters yields,
\begin{align}
    \langle (\Delta v_z )^2 \rangle_{V_{\rm rel}} = \mu^2 \langle \Delta v_{m,||}^2 \rangle_{V_{\rm rel}} +  \left(\frac{1-\mu^2}{2}\right) \langle\Delta v_{m,\perp}^2 \rangle_{V_{\rm rel}},
\end{align}
where
\begin{align}
    \langle \Delta v_{m,||}^2 \rangle_{V_{\rm rel}} =&\frac{4\pi G^2 M_c^2 n_c \Delta t}{V_{\rm rel}} \frac{\Lambda^2}{1+\Lambda^2} \\
    \langle\Delta v_{m,\perp}^2 \rangle_{V_{\rm rel}} =& \frac{4\pi G^2 M_c^2 n_c \Delta t}{V_{\rm rel}} \left[ \ln \left ( 1 +\Lambda^2 \right) - \frac{\Lambda^2}{1+\Lambda^2}  \right] 
\end{align}
where $\Lambda = \Lambda(V_{\rm rel})$ as before. Observe, second moments are non-zero for both parallel and perpendicular contributions. 
\par
We proceed as for the first moment, and integrate over all possible relative velocities,
Finally, the average over cloud velocities can be written as an integral over the three-dimensional velocity distribution,
\begin{align}
\left\langle (\Delta v_z)^2 \right\rangle=\int \mathrm{d}^3 \mathbf V_{\rm c}\,F_c(\mathbf V_{\rm c})\left[\mu^2\left\langle \Delta v_{m,\parallel}^2\right\rangle_{V_{\rm rel}}
+\frac{1-\mu^2}{2}\left\langle \Delta v_{m,\perp}^2 \right\rangle_{V_{\rm rel}}\right].
\label{eq:dv_sq-exact}
\end{align}
The expressions above are written for a fixed stellar velocity $\mathbf v_\star$. The moments used in the one-dimensional simulations are additionally averaged over the stellar in-plane velocity distribution, \(\int \mathrm d^2\mathbf v_\perp\,g(\mathbf v_\perp|\tau)\).

\subsection{Approximations of velocity moments}
To evaluate the exact velocity moments in equations (\ref{eq: dv-exact}) and (\ref{eq:dv_sq-exact}), we adopt a Monte Carlo approach as in Appendix \ref{app: relative speed}. Namely, the integral in the moments is evaluated in a similar way to that of equation (\ref{eq:v_rel_integral}). Since there is no analytic solution of these integrals, we must search for a less costly way to evaluate these moments in the test-particle simulation. 
\par
We begin with a crude approximation,
\begin{equation}
    \langle \Delta v_z \rangle \approx -2\pi G^2 M_c^2 n_c \Delta t  \ln \left [ 1 +\Lambda^2 \right] \frac{v_z}{V_{\rm rms}^3}
\end{equation}
where $\Lambda = \Lambda(V_{\rm rms})$ and $V_{\rm rms}$ is defined in equation (\ref{eq:v_rms}). Again, the RMS approximation vastly decreased the first moment at small $v_z$, due to the $V_{\rm rel}^{-3}$ dependence. Analogously to Appendix \ref{app: relative speed}, we add an empirical correction factor, of the form
\begin{align} \label{eq:C_dv}
    C_1(v_z,\tau)=1+\frac{A(\tau)}{1+\left[v_z/W(\tau)\right]^\gamma}
\end{align}
where $A(\tau)=a_0+a_1\exp(-\tau/\tau_s)+a_2\tau+a_\infty$, and $W(\tau)=w_0+w_1[\sigma_R^2(\tau)+\sigma_\phi^2(\tau)+3\sigma^2]^{1/2}$. Here, $\sigma$ is the 1D velocity dispersion of the clouds. The parameters are fitted to the ratio of the exact moment to the RMS approximation over the $(v_z,\tau)$ grid and then interpolated to get point-wise corrections.
\par
We proceed similarly for the second moment, beginning with an approximation,
\begin{align}
    \langle (\Delta v_z )^2 \rangle \approx & \frac{4\pi G^2 M_c^2 n_c \Delta t}{V_{\rm rms}} \times \nonumber \\
    &\left[ \frac{v_z^2}{V_{\rm rms}^2 }\frac{\Lambda^2}{1+\Lambda^2} + \frac12\left( 1 -  \frac{v_z^2}{V_{\rm rms}^2} \right  ) \left[ \ln \left ( 1 +\Lambda^2 \right) - \frac{\Lambda^2}{1+\Lambda^2}  \right] 
    \right],
\end{align}
where $\Lambda = \Lambda(V_{\rm rms})$ as before. The correction is empirically chosen as a function of velocity only,
\begin{align}
    C_2 (v_z) = 1 + & a_0 \exp[-a_1(v_z-a_2)^2] - w_0 \exp[-(v_z-w_1)^2/(2w_2^2)],
\end{align}
but the fitted parameters are allowed to vary with age. Similarly to the first moment, we fit on a grid in $(v_z,\tau)$ and then interpolate to get pointwise corrections. 

\subsubsection{Sources of error} \label{app:sources of error}
\par
The dominant source of error is replacing the full integral over relative velocities by an RMS relative speed. This approximation loses information about the angular structure of the cloud velocity distribution, especially at small $v_z$. In particular, for the first moment, the dominant $V_{\rm rel}^{-3}$ dependence in the integral, for low $v_z$, induces a bias, since $\langle V_{\rm rel}^{-3}\rangle \geq  \langle V_{\rm rel}\rangle^{-3}$. Despite accurate approximations for $\langle V_{\rm rel} \rangle \sim V_{\rm rms}$, it is less accurate for moments of higher powers of $V_{\rm rel}$. Hence, the RMS approximation underestimates the magnitude of the drag, and thus our fitted corrections, $C_1 > 1$ are greater than 1. Moreover, the exact first moment must go to $0$ as $v_z \rightarrow 0$ by symmetry, but numerically there might be small Monte Carlo variance due to finite samples, inducing a small asymmetry which is why the fractional error may blow up here, even though the absolute error is negligible. We therefore conclude that the corrected approximation is accurate over the full velocity range used in the simulations, with fractional errors typically within $\sim 3\%$ for the first moment. The residual asymptotic behaviour of the error at very small $v_z$ is numerical, arising from the difficulty of evaluating the exact moment where both the moment itself and the relative-velocity denominator become small. The raw approximation for the second moment is also underestimated at small $v_z$, for similar reasons, although the bias is weaker because the leading dependence is closer to $V_{\rm rel}^{-1}$ rather than $V_{\rm rel}^{-3}$. After applying the empirical correction, the second moment is recovered to the $\sim 1\%$ level.
\par
There are two additional nonlinearities that contribute to the error between the exact and approximated moments. Firstly, the Coulomb factor, $\ln(1+\Lambda^2)$, depends on speed through $b_{\rm 90}$ (see equation (\ref{eq:b90})). Secondly, the projection onto $v_z$, via $\mu$ couples the orientation to the total relative speed and there is covariance between $\mu$ and the Coulomb factor, which is neglected when using the RMS approximation. In practice, we choose $b_{\rm min}$ as in equation (\ref{eq:bmin}). These two considerations induce error in the intermediate $v_z$ region, where $b_{\rm min}$ transitions, although our corrections $C_1$ and $C_2$ are flexible enough to absorb these errors. 
\par
Finally, in the high $v_z$ region, $\Lambda$ approaches a constant, since, $b_{\rm min} \rightarrow r_c$, so both moments become slowly varying functions of $V_{\rm rel}$, and the RMS approximation at large relative velocities becomes a much better proxy for the true $V_{\rm rel}$. The correction factors approach unity, $C_1,C_2 \rightarrow 1$, and the residual tends to zero.


\bsp	
\label{lastpage}
\end{document}